\documentclass[12pt]{iopart}

\usepackage{graphicx,setspace}
\usepackage{dsfont}
\usepackage{amssymb}
\usepackage{mathrsfs}
\usepackage{psfrag}

\begin{document}
\title{Horizon effects with surface waves on moving water}
\author{Germain Rousseaux$^{1}$,  
Philippe Ma\"{i}ssa$^1$, Christian Mathis$^1$, Pierre Coullet $^1$,
Thomas G Philbin$^{2}$
and Ulf Leonhardt$^2$}
\address{$^1$ Universit\'{e} de Nice-Sophia Antipolis, 
Laboratoire J.-A. Dieudonn\'{e},
UMR CNRS-UNS 6621,
Parc Valrose,
06108 Nice Cedex 02, 
France.}
\address{$^2$ School of Physics and Astronomy, University of St Andrews,
North Haugh, St Andrews KY16 9SS,
Scotland, UK.}
\ead{Germain.Rousseaux@unice.fr}

\begin{abstract}
Surface waves on a stationary flow of water are considered, in a linear model that includes the surface tension of the fluid. The resulting gravity-capillary waves experience a rich array of horizon effects when propagating against the flow. In some cases three horizons (points where the group velocity of the wave reverses) exist for waves with a single laboratory frequency. Some of these effects are familiar in fluid mechanics under the name of wave blocking, but other aspects, in particular waves with negative co-moving frequency and the Hawking effect, were overlooked until surface waves were investigated as examples of analogue gravity [Sch\"utzhold R and Unruh W G 2002 {\it Phys. Rev.} D  {\bf 66} 044019]. A comprehensive presentation of the various horizon effects for gravity-capillary waves is given, with emphasis on the deep water/short wavelength case $kh\gg 1$ where many analytical results can be derived. A similarity of the state space of the waves to that of a thermodynamic system is pointed out.
\end{abstract}


\section{Introduction}
The interest in black-hole analogues has been mainly driven by the intriguing possibility of observing Hawking radiation in the laboratory~\cite{Unruh,Novello,Volovik,BLV,SUbook,Fiber}. In addition to the experimental challenges, this pursuit has important theory implications because of the well-known weakness in the derivation of the Hawking effect for real black holes~\cite{tHooft,Jacobson}. Hawking's semi-classical calculation~\cite{Hawking} is based on a consideration of fields that are assumed to have no appreciable gravitational effect compared to the black hole, but the derivation contradicts this assumption because the fields attain arbitrarily high frequencies (and therefore energies) at the horizon. This so-called trans-Planckian problem reveals the lack of any proper understanding of quantum-gravitational effects. Unfortunately, the question of whether black holes really radiate does not seem to be amenable to experimental investigation because such radiation would be completely swamped by the cosmic microwave background. In black-hole analogues the trans-Planckian problem is avoided by dispersion that limits the blue-shifting of waves at horizons~\cite{Unruh95}. The Hawking effect has been found to persist even in the presence of dispersion~\cite{Unruh95,Brout,Corley,Jacobson99,Himemoto,Saida,Unruh05,SU08}, although in general numerical simulations must be resorted to in order to verify this. The lack of a good understanding of the Hawking effect with arbitrary dispersion is an important and presumably solvable problem, whereas a proper understanding of the Hawking effect in real black holes requires unknown physics to deal with the infinite blue-shifting. Analogue systems therefore provide the opportunity to understand the Hawking effect using known physics and to verify it experimentally; neither of these tasks can be achieved with real black holes.

The introduction of dispersion into the Hawking effect opens up a variety of new possibilities. Indeed, one of the lessons of analogue systems is that the physics of horizons, including the Hawking effect, has a breadth and richness that is not immediately apparent from the case of real black holes. One can consider white-hole and black-hole horizons~\cite{SU,Fiber}, that both give rise to the Hawking effect; horizons allow the blue-shifting and red-shifting of probe waves~\cite{Fiber}; depending on the dispersion, the probe waves can bounce back at horizons, or go straight through, or bounce back and forth a number of times (see below); two horizons can communicate leading to runaway quantum Hawking radiation or a classical instability~\cite{BHlaser,SU,SUbook,Barcelo}.

\begin{figure}[!htbp]
\begin{center}
\includegraphics[width=12cm]{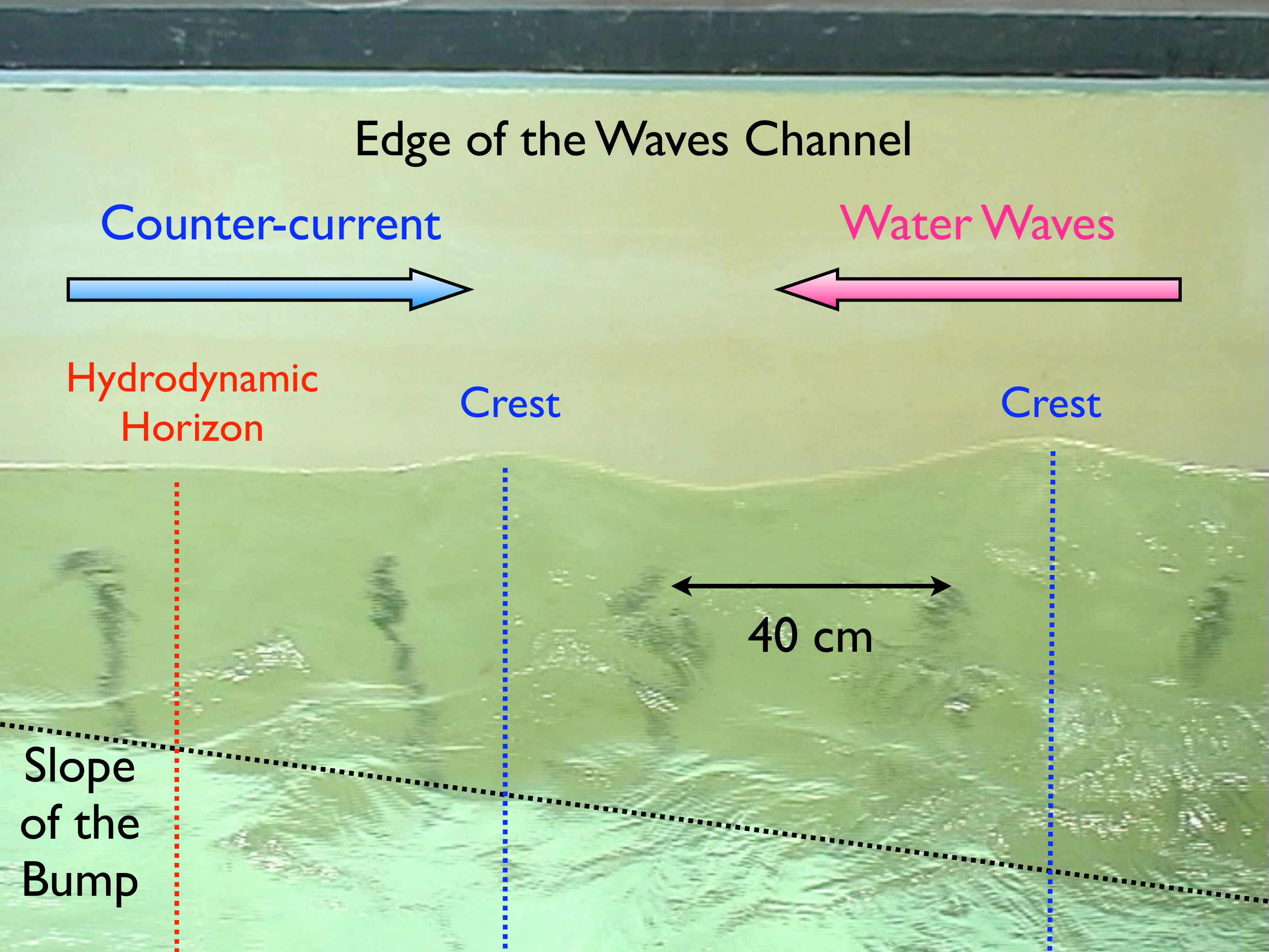}
\caption{Experimental white-hole horizon in hydrodynamics. The surface waves propagate against the flow up to the point where the flow speed matches the group velocity of the waves. The flow speed is higher on left than on the right because of the slope on the bottom of the tank.}
\label{horizoninfluid}
\end{center}
\end{figure}

\begin{figure}[!htbp]
\begin{center}
\includegraphics[width=12cm]{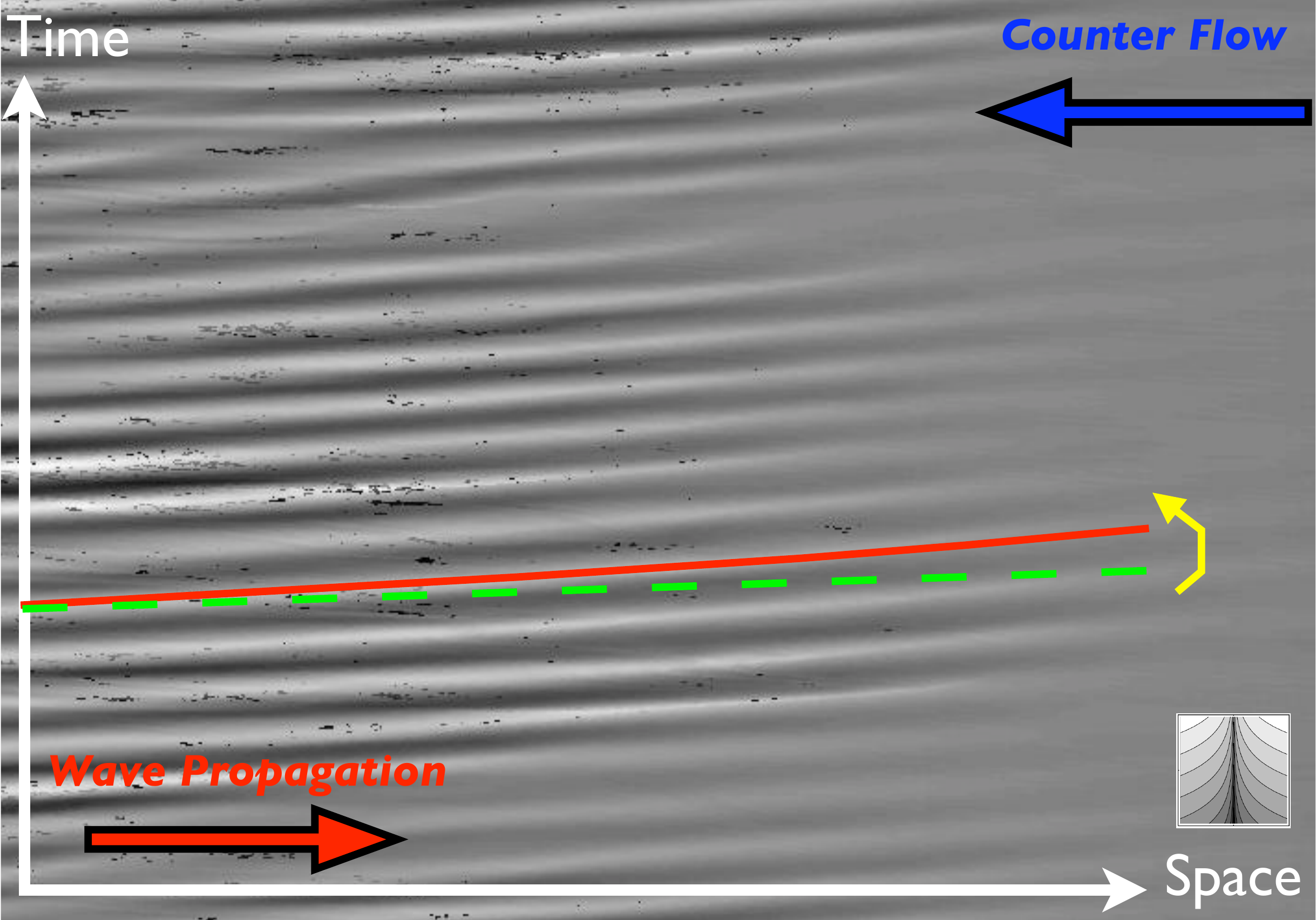}
\caption{Space-time diagram of a plane wave encountering a white-hole horizon. In this diagram the wave propagates to the right on a flow moving to the left. Grey values denote the water height and show clearly the world-lines of the crests and troughs. The green dotted line shows the initial slope of the world-line of the incoming wave, which is inversely proportional to the phase velocity. The red line is the world-line of a crest, the slope of which increases as the wave reaches the horizon. The phase velocity thus decreases at the horizon, leading to a decrease of the wavelength, a blue-shifting that is limited by dispersion. The inset shows the behaviour of rays at a non-dispersive white-hole horizon, where there would be infinite blue-shifting of waves (rays originating on both sides of the horizon are shown).}
\label{blueshifting}
\end{center}
\end{figure}

Perhaps the least exotic black-hole analogue that has been proposed is that of surface waves on a moving fluid. Sch\"utzhold and Unruh~\cite{SU} showed that long-wavelength surface waves in a shallow moving fluid obey the Klein-Gordon equation in a curved space-time geometry. By varying the flow speed so that it exceeds the wave group velocity in some region, one produces a horizon. Close to the horizon the long-wavelength assumption breaks down and the detailed behaviour is governed by the dispersion; the wave evolution is no longer described by an effective space-time metric and there is therefore no trans-Planckian problem of infinite blue-shifting. Further results for surface waves from an analogue-gravity perspective were given in~\cite{visser}. Recently, we performed experiments to investigate the behaviour of surface waves at a white-hole horizon, using water waves in a wave tank with a counter-flow~\cite{NJP}. The speed of the counter-flow was varied along the tank by the presence of a sloping region on the bottom. Figure~\ref{horizoninfluid} shows how waves propagating against the flow are blocked at the point where the counter-current reaches the group velocity of the wave; this blocking line is the white-hole horizon. Video of the incoming wave was used to produce the space-time diagram in Figure~\ref{blueshifting} (note that the directions of the flow and the wave are reversed in the plot of Figure~\ref{blueshifting} compared to Figure~\ref{horizoninfluid}). The linear features with positive slope in the space-time diagram are the evolution of the wave crests and troughs in time---the world-lines of the crests and troughs. The inverse $\rmd x/\rmd t$ of the slope of the world-lines gives the speed of the crests and troughs, the phase velocity $\omega/k$ of the wave since the phase is $\int(k\rmd x-\omega\rmd t)$; the lines curve upward as the wave reaches the white-hole horizon, showing a decrease in the phase velocity. The wavelength in the laboratory is revealed by drawing a horizontal line in the diagram and measuring the distance between two crests; because of the curving upward of the crest world-lines, the wavelength is seen to decrease as the horizon is approached. This is the characteristic wavelength-shortening (``blue-shifting") of waves at a white-hole horizon, also observed in an optical analogue~\cite{Fiber}. The inset in Figure~\ref{blueshifting} shows the behaviour of rays at a white-hole horizon where there is no dispersion; the rays stick at the horizon which corresponds to an infinite blue-shifting of the waves (the trans-Planckian problem). It is the dispersion of the surface waves that limits the amount of blue-shifting at the horizon; in this paper we will discuss some of the rich set of possible behaviours.

Not surprisingly, the blocking of waves by a counter-flow that exceeds the wave group velocity is a phenomenon that is well-known in the fluid-mechanics community~\cite{FS,Dingemans,Peregrine,Suastika,Igor}. The blue-shifting of waves at the blocking line (white-hole horizon) is also well-known and has been investigated experimentally~\cite{Chawla1, Chawla2,Suastika}. There is however another possible process in the interaction of a wave with a counter-flow, one that does not seem to have been considered by the fluid-mechanics community, even though it is a possibility clearly visible in the dispersion relation~\cite{SU,NJP}: this process is the Hawking effect. In a stationary flow, where the flow speed at each point in the tank is constant in time but varies from point to point, the frequency in the laboratory frame of a surface wave on the flow is conserved. As we have seen, the wavelength in the laboratory is not conserved, so a white-hole horizon converts the wavelength of a wave propagating against the flow to a different wavelength while conserving its frequency in the laboratory. Figure~\ref{blueshifting} is an example of such a wave; it is right-moving against the flow (ingoing against the flow) with positive (angular) frequency $\omega$ and positive wave number $k$ and the horizon increases $k$ while keeping $\omega$ fixed. An important quantity is the frequency of the wave in a frame co-moving with the flow; this co-moving frequency is not conserved, but the ingoing wave and the blue-shifted wave with higher $k$ both have positive co-moving frequencies~\cite{SU,NJP}. It turns out that there is often a solution of the dispersion relation for the fixed positive input frequency $\omega$ that has a \emph{negative} $k$ and a \emph{negative} co-moving frequency~\cite{SU,NJP}. The laboratory frequency $\omega$ of the input wave must be conserved in the wave evolution but when there is a wave with negative co-moving frequency at same value of $\omega$, there exists the possibility that it could be generated in the interaction of the input wave with the counter-flow. These waves with negative co-moving frequency would be produced in addition to the blue-shifted waves with positive co-moving frequency; this process is the Hawking effect, which is at root a classical effect, although with extraordinary quantum implications~\cite{Hawking,Unruh76}. The strangest feature of the Hawking effect is that it is an amplification of the ingoing wave, an extraction of energy from the flow (or from whatever background creates the horizon). The famous quantum Hawking radiation is a rather straightforward consequence of the classical Hawking effect when the fields underlying the waves are quantized (the role of input waves is then played by the quantum vacuum and the energy extraction in the Hawking process allows the spontaneous creation of field quanta)~\cite{Hawking}.

In our experiments described above~\cite{NJP} we observed indications of waves with negative co-moving frequency. Numerical simulations indicated that the Hawking effect would be unobservable in the regimes covered in the experiments~\cite{NJP}, but it is not clear how well the wave evolution was described by the theoretical model used in the simulations. It appears that our paper was the first experimental search for the Hawking effect with surface waves, and further experiments are needed. We hope through this paper to increase awareness of the Hawking effect among researchers in fluid mechanics by showing how it has been overlooked in the existing literature on wave blocking. Complementary to this, we hope to show those familiar with horizon physics some of the rich horizon effects that occur in models of wave blocking that are used in fluid mechanics. Such surprising connections between apparently disparate areas of physics (and engineering) are often a source of inspiration to both sides.

It is important to stress that the model of surface waves considered in this paper is a linear one: the dynamics is given by a linear wave equation, but with a nonlinear dispersion relation. As is well known, surface waves behave non-linearly if the wave amplitude is large enough, and this presents a potential problem since the shortening of the wavelength at a horizon necessarily leads to an increase in the wave amplitude. The limiting of blue-shifting by dispersion also limits the amplitude increase at blocking lines and, for the linear model to be applicable, it is important that the wave amplitude not get too large. The maximum amplitude attained during the interaction with a counter-flow can to some extent be controlled through the amplitude of the incident wave, and dissipation (see next paragraph) also helps to limit the amplitude. The interaction of nonlinear waves with a counter-flow is a highly complex process that is not discussed here (see for example~\cite{Choi}). 

Also neglected in the model used below is the viscosity of the fluid, and this limits the accuracy of the results when wavelengths are blue-shifted down to very small (capillary) wavelengths since viscosity quickly damps such waves. But it is easy to make allowance for this latter drawback of the model by bearing in mind the limited propagation length of capillary waves.

\section{Surface waves on a stationary flow}
The subjects of black holes and water waves have a historical connection through the figure of Pierre-Simon de Laplace. In his  "Exposition du Syst\`eme du Monde" in 1796, Laplace famously introduced the term {\it \'etoile sombre} (dark star) to denote an object whose gravitational field is strong enough to prevent light from escaping (the same concept had been described in 1783 by the Reverend John Michell in a letter to Cavendish)~\cite{Schaffer}. Laplace is also well known for having derived (in 1775) a dispersion relation for surface waves on water~\cite{Darrigol}. When the water has a background flow (which does not vary with depth) and the effect of surface tension is included, the dispersion relation takes the form widely used in fluid mechanics to describe waves on moving water~\cite{FS,Badulin,Dingemans,Huang}:
\begin{equation}  \label{disprel}
(\omega -Uk)^2 = \left( gk+\frac{\gamma}{\rho} k^3 \right) \tanh(kh).
\end{equation}
Here $\omega$ is the (angular) frequency of the wave in the laboratory frame and $k$ the wave-number; $U$ is the speed of the flow, $h$ the depth of the fluid, $g$ the gravitational constant, $\rho$ the fluid density and $\gamma$ the surface tension. For water, $\rho=1000\,\mathrm{kg}\,\mathrm{m}^{-3}$ and $\gamma=0.073\,\mathrm{N}\,\mathrm{m}^{-1}$. Equation (\ref{disprel}) is a one-dimensional dispersion relation suitable for water-tank experiments. When the flow is stationary ($U$ independent of time, but gradually varying in space), $\omega$ is a constant but $k$ varies with spatial position  $x$. Waves described by (\ref{disprel}) are a consequence of gravity ($g$) and surface tension ($\gamma$) and are called gravity-capillary waves.  For small $k$ the gravity term dominates, which we call the gravity regime, whereas for large $k$ surface tension dominates, giving the capillary regime. The pure-gravity case corresponds to $\gamma=0$.

The quantity $\omega-Uk$ is the frequency in a frame co-moving with the fluid. Hence the positive, respectively negative, square roots of (\ref{disprel})
\begin{equation}  \label{disprel2}
\omega -Uk= \pm\sqrt{\left( gk+\frac{\gamma}{\rho} k^3 \right) \tanh(kh)}
\end{equation}
correspond to positive, respectively negative, co-moving frequencies. As described in the Introduction, the Hawking effect is the generation of a wave on the negative branch of (\ref{disprel2}) from a wave on the positive branch, through interaction with a counter-flow. It is a remarkable fact that in the extensive fluid-mechanics literature on the waves (\ref{disprel}), including the pure-gravity case $\gamma=0$, there seems never to have been any consideration of the possibility of conversion of waves from the positive to the negative branch of (\ref{disprel2}) through the blocking effect~\cite{FS,Badulin,Dingemans,Peregrine,Smith,PS,Basovich,Chawla1,Chawla2,Suastika,Igor,Baschek}. If such a conversion process had been investigated, the Hawking effect would presumably have been (re-)discovered in fluid mechanics. 

\begin{figure}[!htbp]
\begin{center}
\includegraphics[width=12cm]{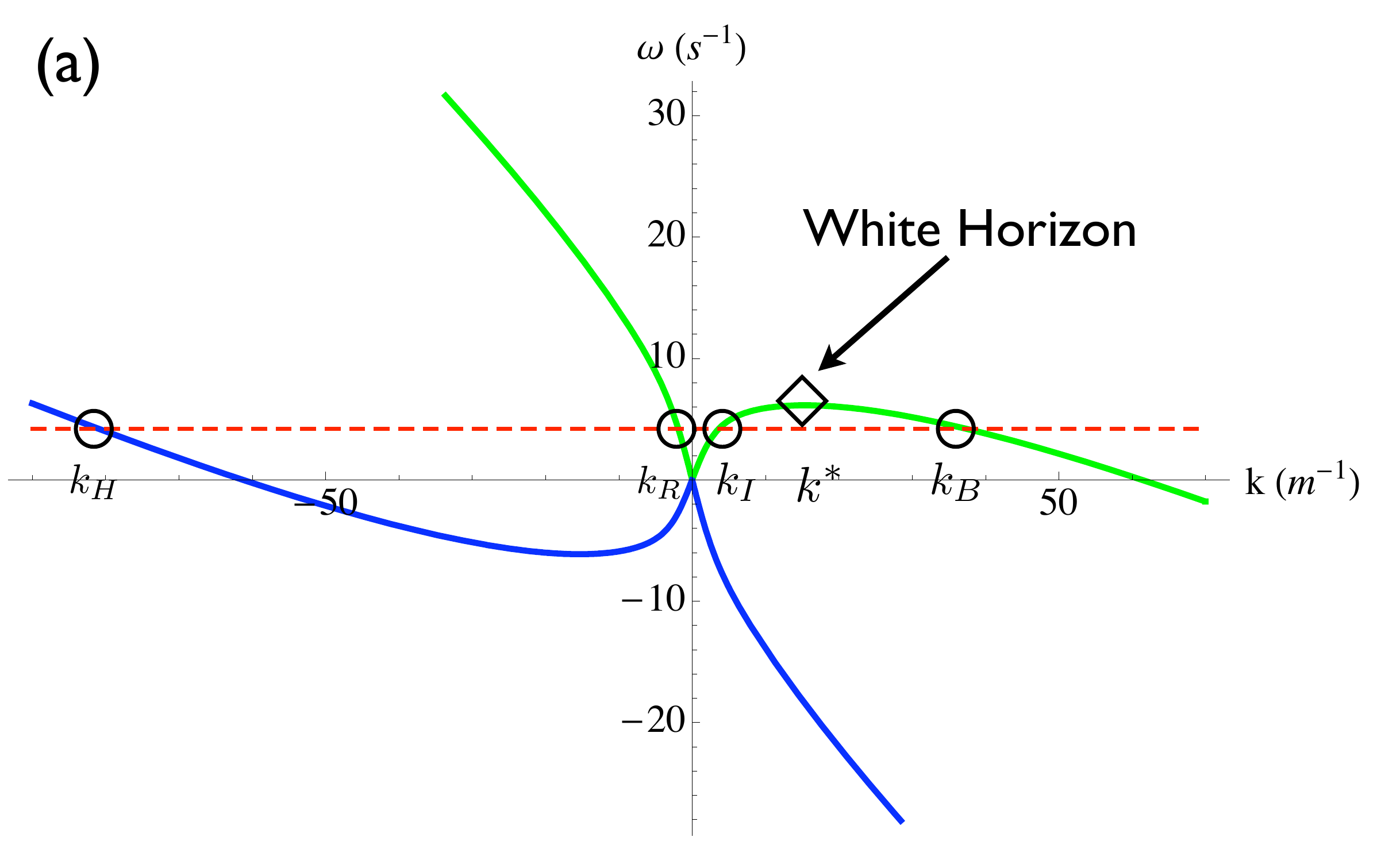}
\includegraphics[width=12cm]{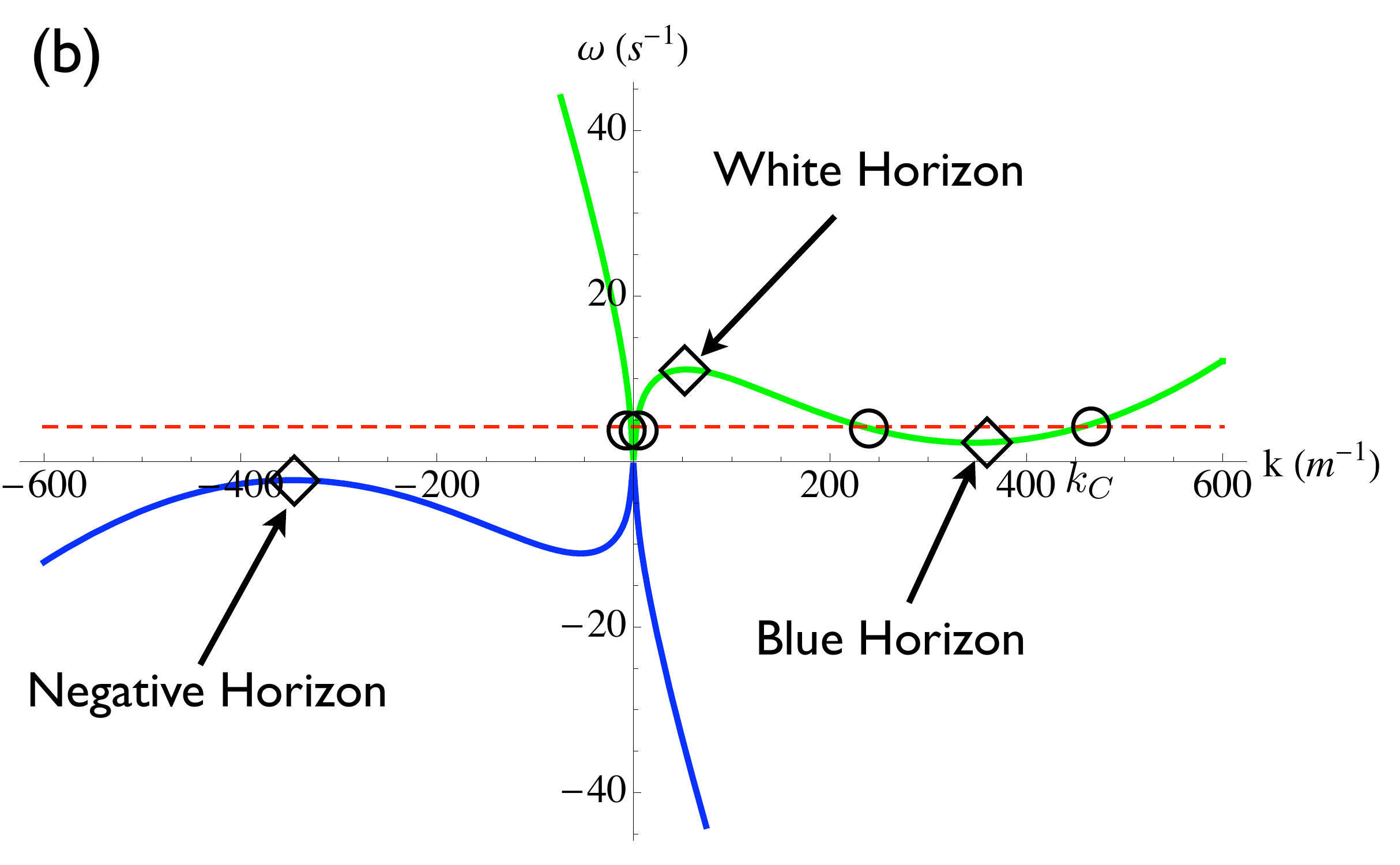}
\caption{Dispersion relation for surface waves propagating on a water flow with a given velocity $U<0$; (a) shows the pure gravity case ($\gamma=0$) and (b) shows the gravity-capillary case. The green curves are the positive branch of the right-hand side of (\ref{disprel3}) (positive co-moving frequency) and the blue curves are the negative branch (negative co-moving frequency). Intersections of these curves with the horizontal red line (value of $\omega$) show possible waves at that $\omega$. If $U$ becomes more negative the green and blue curves rotate clockwise about the origin. Local maxima and minima of the green and blue curves show the possibility of blocking waves by means of appropriate velocity profiles $U(x)$ (see Figure~\ref{rays}). A water depth $h=0.4\,\mathrm{m}$ is assumed. }
\label{dispersion}
\end{center}
\end{figure}

Solutions of the dispersion relation (\ref{disprel}) are usually represented graphically. We obtain from (\ref{disprel2})
\begin{equation}  \label{disprel3}
\omega =Uk \pm\sqrt{\left( gk+\frac{\gamma}{\rho} k^3 \right) \tanh(kh)}\,.
\end{equation}
Figure~\ref{dispersion} plots both branches of the right-hand side of (\ref{disprel3}) as functions of $k$, for a fixed value of $U<0$; figure (a) shows the pure-gravity case ($\gamma=0$) and figure (b) shows the full dispersion relation with surface tension included. The positive branch of (\ref{disprel3}), corresponding to positive co-moving frequency, is shown in green while the negative branch, corresponding to negative co-moving frequency, is shown in blue. The intersection of these curves with a given horizontal line, such as the red line in the figures, gives the possible waves for the frequency $\omega$ given by that line. In a stationary flow $\omega$ is conserved but the plots of the right-hand side of (\ref{disprel3}) change with spatial position as $U(x)$ changes and one can trace the evolution of a given solution by following its intersection point with the horizontal line of fixed $\omega$. As one traces the evolution of the intersection point, the changing group velocity $\rmd\omega/\rmd k$ is the slope of the tangent to the curve at the point of intersection with the horizontal line. 

\begin{figure}[!htbp]
\begin{center} 
\includegraphics[width=12cm]{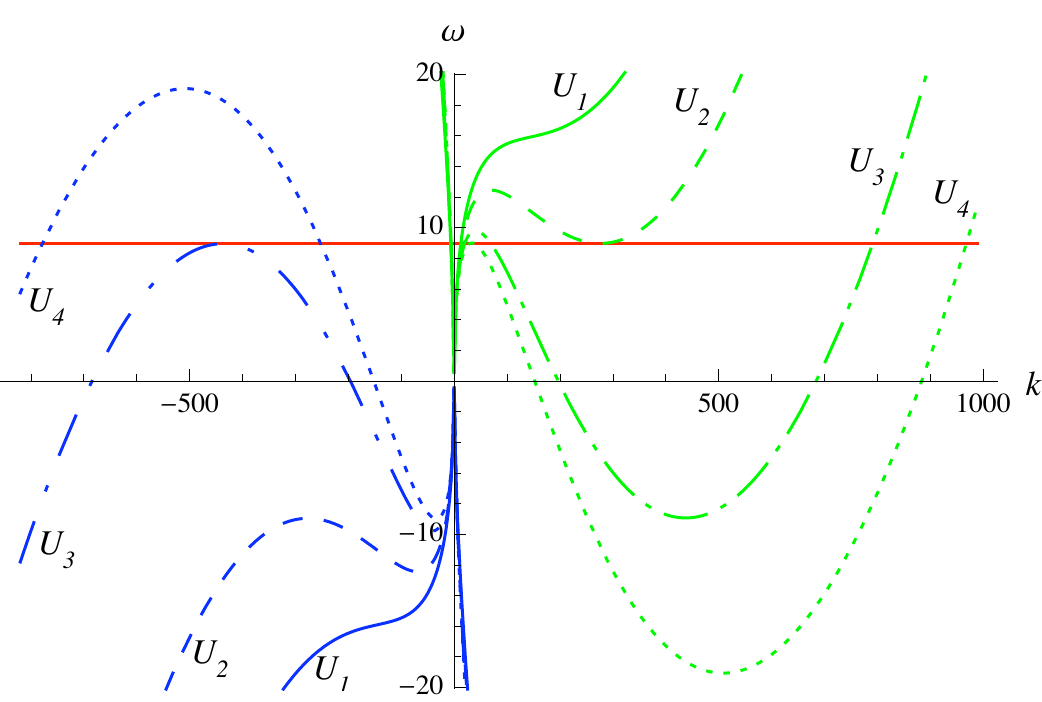} 

\vspace{5mm}

\includegraphics[width=10cm]{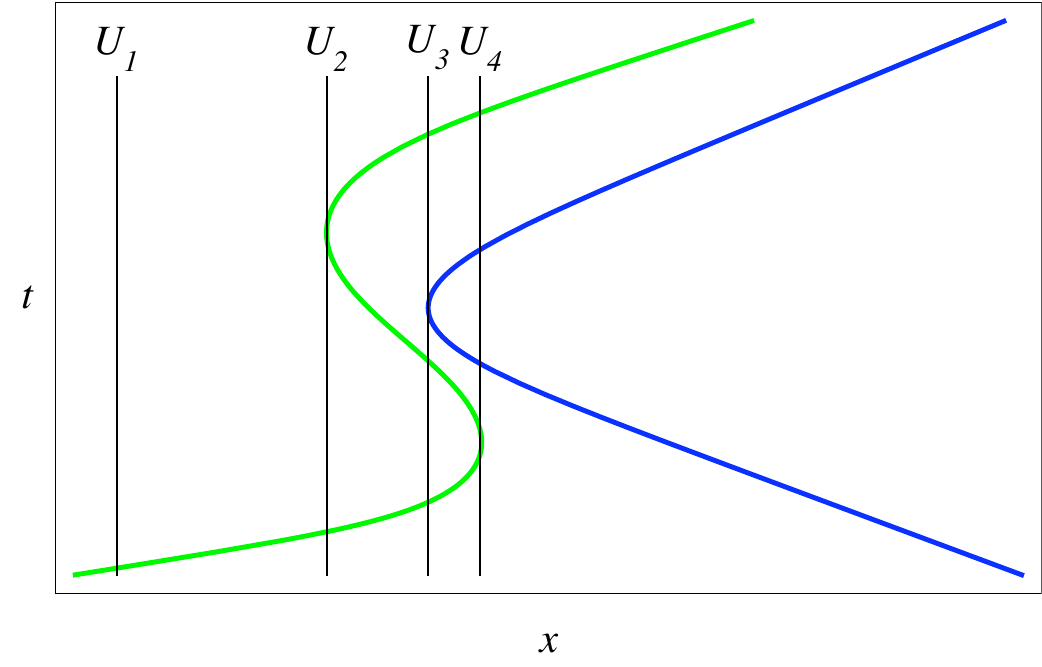}
\caption{Graphical solution and numerical ray solution for gravity-capillary waves on a water counter-flow. The green/blue curves refer to positive/negative co-moving frequency. Four values of the velocity profile $U(x)<0$ are shown in the dispersion plots (upper) and the $x$-positions where the profile takes these four values are shown in the ray solutions (lower). The values are (in $\mathrm{m}\,\mathrm{s}^{-1}$) $U_1=-0.17$, $U_2=-0.2035$, $U_3=-0.2536$ and $U_4=-0.275$. We use a hyperbolic tangent function $\tanh ax$ to describe the variation of the velocity profile, with a typical length $a=0.5\,\mathrm{m}$. The conserved frequency $\omega$ of the waves (red horizontal line in the dispersion plots) corresponds to a period $T=0.7\,\mathrm{s}$.  A water depth $h=0.4\,\mathrm{m}$ is assumed.}
\label{rays}
\end{center}
\end{figure}

In the pure-gravity case (Figure~\ref{dispersion}~(a)) there are at most four real roots of the dispersion relation ($k_I$, $k_B$, $k_R$ and $k_H$). The solution $k_I$ is a right-moving wave in the laboratory, having positive phase and group velocities, propagating against a left-moving counter-flow $U<0$. If the flow speed $|U(x)|$ increases as the wave moves to the right ($U$ becomes more negative) the green and blue curves tip over clockwise about the origin (as in Figure~\ref{rays}), so the root $k_I$ increases---the wave is blue-shifted. When the wave reaches a point where the flow speed has increased to make the roots $k_I$ and $k_B$ coalesce at a local maximum of the green curve, the group velocity of the wave is zero---it has been blocked at a white-hole horizon. The wave has been stopped by a negative ``group acceleration" that is still non-zero at the blocking point so the group velocity decreases to negative values; the wave moves back to the left in the laboratory on the $k_B$ root of the dispersion relation, back into the region where the counter-flow is slower than the blocking speed and where the wave previously had wave number $k_I$. The ingoing wave $k_I$ has thus been blue-shifted to $k_B$ by the white-hole horizon; the blue-shifted wave $k_B$ has positive \emph{phase} velocity, so its crests move to the right in the laboratory, but it has negative group velocity. The third real root of the dispersion relation with positive co-moving frequency (green curve) in 
Figure~\ref{dispersion}~(a) is $k_R$; this is simply a wave propagating in the same direction as the flow, to the left with negative phase and group velocities. The solution $k_R$ is rather trivial and is of no interest for horizon effects. 

The root $k_H$  in Figure~\ref{dispersion}~(a) has negative co-moving frequency (blue curve) and is of great interest for horizon effects; the conversion of some of the input wave $k_I$ to $k_H$ is the Hawking effect. Since it has a negative wave number, the wave $k_H$ has a negative phase velocity in the laboratory; its crests move backwards relative to the direction of the ingoing wave, in contrast to $k_I$ and $k_B$, which both have positive phase velocity. It is essential to understand that the three waves $k_I$, $k_B$ and $k_H$ are all propagating to the right \emph{relative to the fluid}, even though in the laboratory $k_H$ has a phase velocity pointing left and both $k_B$ and $k_H$ have a group velocity pointing left ($k_R$ is the root corresponding to a wave moving to the left relative to the fluid). Unlike the blue-shifting of $k_I$ to $k_B$, the existence of conversion from $k_I$ to $k_H$ cannot be deduced from dispersion plots, which only reveal it as a possibility. The amount of conversion of $k_I$ to $k_H$ depends on the details of the dispersion and the velocity profile $U(x)$. In simple cases involving limited dispersion the Hawking effect is determined by the slope $\rmd U(x)/\rmd x$ of $U$ at the horizon (this slope is the analogue of the surface gravity of a black hole, the acceleration due to gravity at the horizon), but for general dispersion and velocity profiles no analytical formula for the size of the effect has been found and one must resort to numerical simulations of the wave evolution. One aspect of the challenge to find a good intuitive understanding of the Hawking effect is apparent from the description of the horizon given above: this was taken as the point where the flow speed matched the group velocity of the blue-shifting wave. But the phase velocity of this wave is greater than its group velocity, so one can have a group-velocity horizon but no phase-velocity horizon. On the other hand, one can have both a group- and a phase-velocity horizon, with in principle an arbitrary distance between these two horizons and a completely different value of $\rmd U(x)/\rmd x$ at each horizon. The size of the Hawking effect is influenced by these and other factors.\footnote{An extension of the standard analytical results to the dispersive case was given in~\cite{SU08} but only for two specific velocity profiles, both of which gave a group- and a phase-velocity horizon.} There is also the further possibility of having the maximum flow speed close to but less than that required for a group-velocity horizon. In this case one would expect some wave tunneling into the blue-shifted root $k_B$, and perhaps also into $k_H$ (tunneling of surface waves has been studied in~\cite{sti79}). Numerical simulations indicate that this method of generating $k_H$ without a group-velocity horizon is mathematically possible for a steep enough velocity profile, but it should not be possible in practice~\cite{NJP}. In the experiments reported in~\cite{NJP} waves with negative phase velocity were observed even in the absence of a white-hole group-velocity horizon, but, as stated in the Introduction, the origin of those waves is not clear. 

The conversion of $k_I$ to $k_B$ discussed above is well known in fluid mechanics, under the name of wave blocking~\cite{FS,Badulin,Dingemans,Peregrine,Smith,PS,Basovich,Chawla1,Chawla2,Suastika,Igor,Baschek}.  The superposition of the $k_I$ and $k_B$ waves has been shown to be describable by an Airy interference pattern~\cite{Nardin,Smith,PS,Basovich}. In contrast, the root $k_H$ in Figure~\ref{dispersion}(a) has been largely neglected by the fluid-mechanics community. Although the graphical representation of the dispersion relation is standard in fluid mechanics, very few authors~\cite {Peregrine,shy90,shy99,Trulsen} plot the negative-$k$ part, in either the pure-gravity or gravity-capillary cases, and the conversion of $k_I$ to $k_H$ appears not to have been considered.

Turning to the full gravity-capillary case, Figure~\ref{dispersion}~(b) shows (for fixed $U<0$) how the surface tension $\gamma$ changes the dispersion relation at large wave numbers compared to the pure-gravity case in Figure~\ref{dispersion}~(a). For the value of $U$ plotted, the positive co-moving frequency curve (green) has a local minimum as well as a local maximum, and this is also the case for the negative co-moving frequency curve (blue), although the local minimum of the latter curve always occurs at negative laboratory frequency $\omega$. If $U$ becomes more negative the curves tip over clockwise about the origin, so that for larger counter-flow speed there exist roots with negative co-moving frequencies at the $\omega$ shown by the red line. Each local maximum or minimum of the green and blue curves reveals the possibility of reversing the group velocity of a wave with an appropriate velocity profile, as in the discussion of the local maximum in the pure-gravity case (Figure~\ref{dispersion}~(a)). The local maximum of the green curve in Figure~\ref{dispersion}~(b) allows the wave blocking and blue shifting of an incident right-moving wave as in the pure-gravity case. We refer to this blocking line as the white horizon (from white hole). But after the white horizon has reversed the group velocity of the incident wave so that it now moves to the left, this blue-shifting wave encounters another blocking line because of the local minimum of the green curve. We refer to this second blocking line for the blue-shifting wave as the blue horizon. At the blue horizon the group velocity reverses once more to become positive so the wave moves to the right again towards the white horizon. This time the (still blue-shifting) wave goes right through the white horizon, so overall the incident wave undergoes a double bounce. Figure~\ref{rays} shows the graphical solution of the dispersion relation at four values $|U_1|<|U_2|<|U_3|<|U_4|$ of a velocity profile $U(x)<0$ and also a numerical solution of the ray equations for the incident right-moving wave (green curve in the lower figure). The $x$-positions where the velocity profile takes the four values used in the dispersion plots are shown in the ray plot. Rays move at the group velocity and so the wave blocking is clear from the ray plots. Also shown is the ray solution for the wave with negative co-moving frequency (blue curve); this wave is initially left-moving in the laboratory but its group velocity is reversed at a blocking line we refer to as the negative horizon (this horizon does not exist in the pure-gravity case---see~\cite{NJP}). Comparing the dispersion and ray plots in Figure~\ref{rays} (and ignoring the co-propagating wave $k_R$ discussed above) one can see how the increasing counter-flow speed as $x$ increases gives, successively, one $k$ root in the profile where $|U|<|U_2|$, three roots in the region where $U$ lies between $U_2$ and $U_3$, five roots in the region where $U$ lies between $U_3$ and $U_4$, and three roots in the region where $|U|>|U_4|$. One can also see how these roots relate to the ray behaviour. (See~\cite{NJP} for ray plots in the pure-gravity case.)

\begin{figure}[!htbp]
\begin{center} 
\includegraphics[width=10cm]{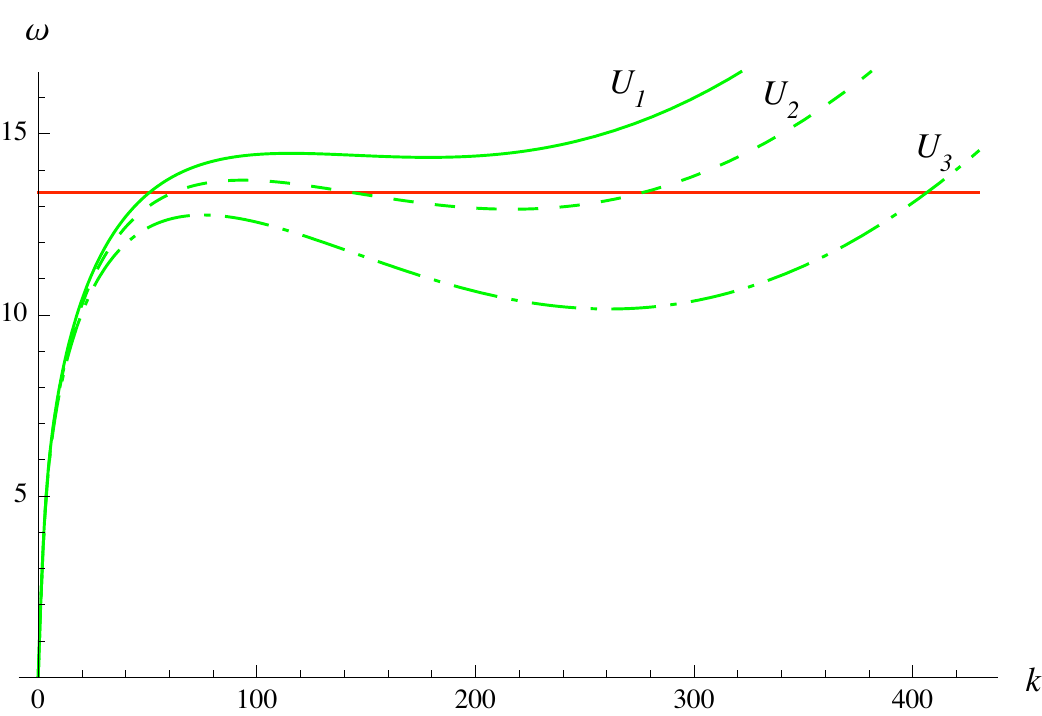} 

\vspace{5mm}

\includegraphics[width=10cm]{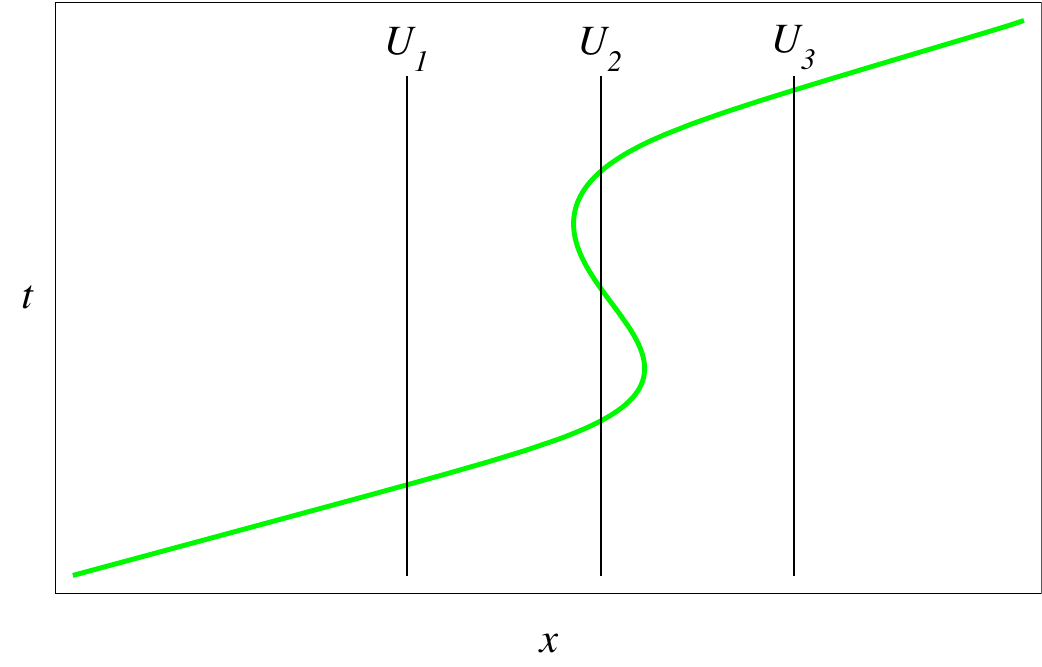}
\caption{Graphical solution and numerical ray solution for gravity-capillary wave with positive co-moving frequency. The conserved frequency $\omega$ corresponds to a period $T=0.47\mathrm{s}$. For this frequency there is no wave with negative co-moving frequency in the range $|U_1|<|U_2|<|U_3|$ of counter-flow speeds used. The three values of the flow velocity shown are (in $\mathrm{m}\,\mathrm{s}^{-1}$) $U_1=-0.1804$, $U_2=-0.1876$, $U_3=-0.1991$.  A water depth $h=0.4\,\mathrm{m}$ is assumed.}
\label{rays2}
\end{center}
\end{figure}

\begin{figure}[!htbp]
\begin{center} 
\includegraphics[width=11cm]{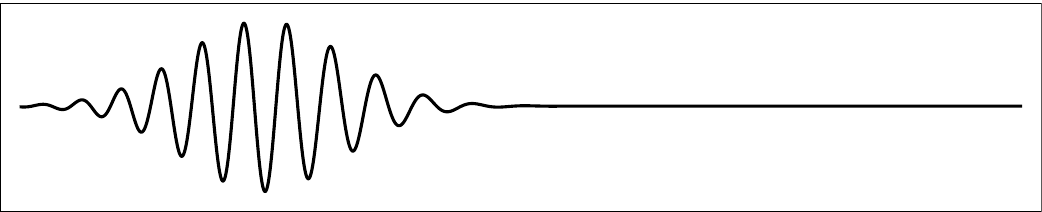} 

\vspace{1mm}

\includegraphics[width=11cm]{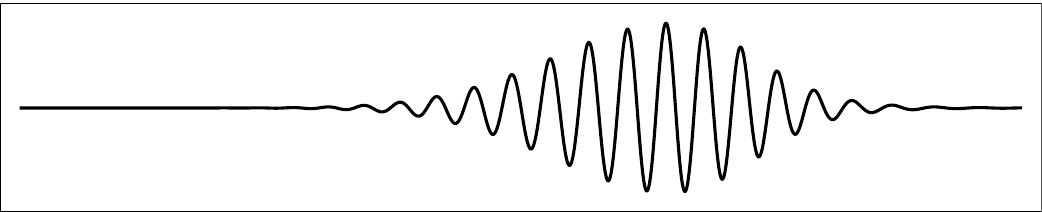}

\vspace{1mm}

\includegraphics[width=11cm]{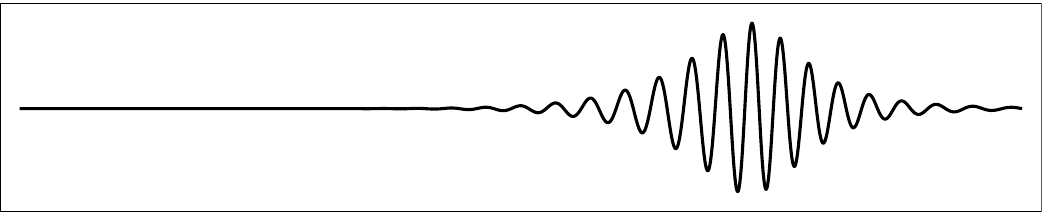}

\vspace{1mm}

\includegraphics[width=11cm]{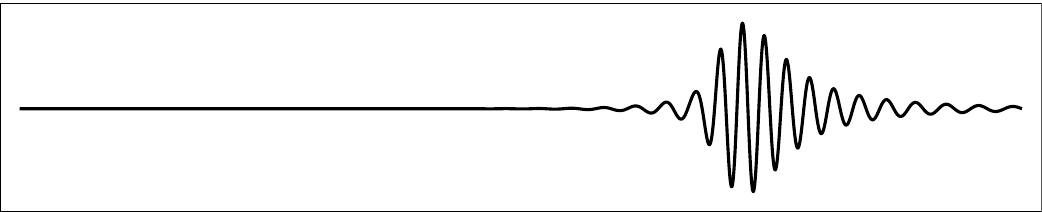}

\vspace{1mm}

\includegraphics[width=11cm]{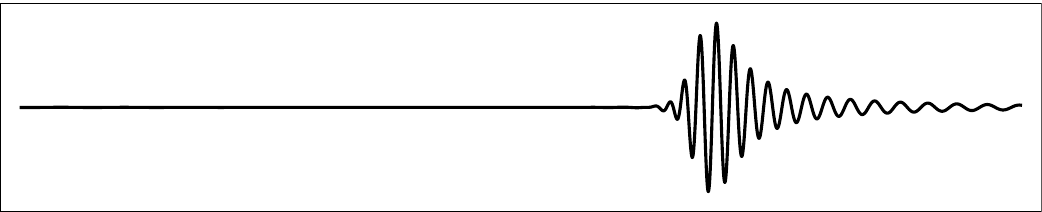}

\vspace{1mm}

\includegraphics[width=11cm]{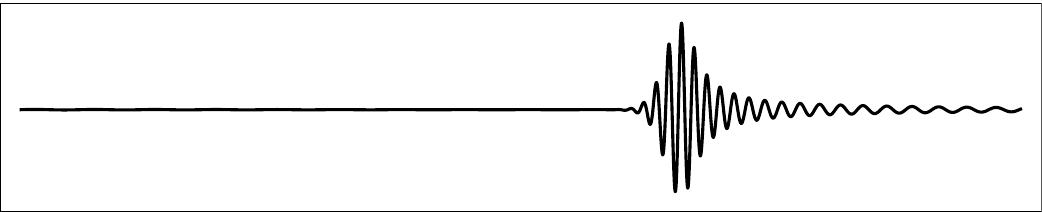}

\vspace{1mm}

\includegraphics[width=11cm]{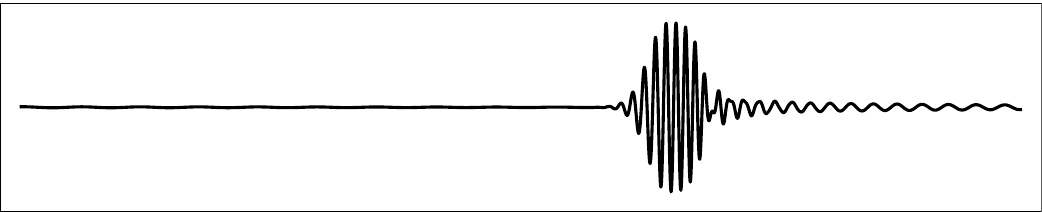}

\vspace{1mm}

\includegraphics[width=11cm]{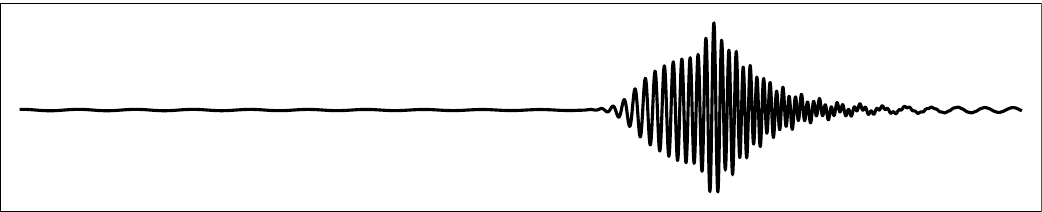}

\vspace{1mm}

\includegraphics[width=11cm]{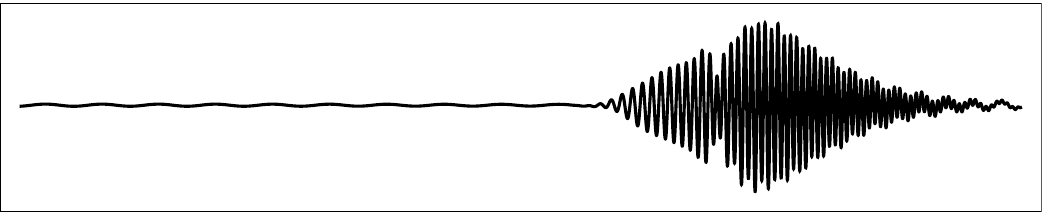}
\caption{Wave packet simulation. The packet is centred on the ray in Figure~\ref{rays2}. }
\label{num}
\end{center}
\end{figure}

Figure~\ref{rays2} shows an example where the frequency $\omega$ and counter-flow profile $U(x)<0$ are such that only a wave with positive co-moving frequency exists. The wave again displays the double-bouncing behaviour. Figure~\ref{num} shows a numerical solution for a wave packet centered on the ray in Figure~\ref{rays2}. This simulation was obtained by solving the scalar wave equation describing the surface wave on the counter-flow; it was shown by Sch\"{u}tzhold and Unruh~\cite{SU} that this equation takes the form of the Klein-Gordon equation in a curved space-time with added higher-order dispersion, in this case the dispersion (\ref{disprel}) of gravity-capillary waves:
\begin{equation}  \label{waveeqn}
(\partial_t + \partial_x U)(\partial_t + U \partial_x) \phi 
= \mathrm{i}\left(g\partial_x-\frac{\gamma}{\rho}\partial^3_x\right) 
\tanh (-\mathrm{i}h\partial_x)\, \phi.
\end{equation}
The method of numerically solving equations of the form (\ref{waveeqn}), for essentially arbitrary dispersion, is described in~\cite{Unruh95}; further examples of this kind of numerical solution for surface waves appear in~\cite{SU} and~\cite{NJP}. In the wave-packet simulation in Figure~\ref{num}, the continuous blue-shifting that accompanies the double bounce is apparent. Because of the spread of frequencies in the wave-packet there is some leakage of the initial wave through the white horizon (first bounce) as well a spreading and separation of frequency components at the blue horizon (second bounce). For extensive numerical simulations of gravity-capillary waves in the presence of a current, see \cite{TM,Trulsen}.

In practice the blue-shifting of incident gravity waves into the capillary regime, as in Figure~\ref{num}, will be limited by viscosity, which is not included in the model we have been discussing. As a consequence, the highly blue-shifted waves produced at the blue horizon will dissipate rapidly. An experimental investigation of these effects was first performed by Badulin {\it et al.}~\cite{Badulin}. These authors observed the initial blocking of waves in the gravity regime (white horizon) and the subsequent conversion into waves in the capillary regime (blue horizon) which then propagated through the original blocking line and vanished through viscous damping~\cite{Badulin}. Gravity-capillary waves on a counter-flow have also been studied experimentally by Klinke and Long~\cite{Klinke}, who produced space-time diagrams of the wave evolution. Trulsen and Mei~\cite{TM,Trulsen} give a theoretical treatment that includes numerical simulations. A recent theoretical survey is given by Huang~\cite{Huang}.  

We have so far discussed only certain features of the dispersion relation (\ref{disprel}). In the next Section we classify in more detail how the presence or absence of the various horizons, and their positions, depend on the conserved frequency $\omega$ and velocity profile $U(x)$.

\section{Results for wavelengths less than the water depth}
The influence of the water depth $h$ in the dispersion relation (\ref{disprel}) disappears when $|k|h\gg 1$, so that $\tanh kh\approx \pm1$ (depending on the sign of $k$). 
This is the case of wavelengths short compared to the water depth and it gives the polynomial dispersion relation 
\begin{equation}  \label{dispshort}
(\omega -Uk)^2 = \pm \left(gk+\frac{\gamma}{\rho} k^3 \right) \qquad \mbox{($+$ for $k>0$,\ \ $-$ for $k<0$)}
\end{equation}
that is easier to handle analytically than the original (\ref{disprel}). The waves considered in Figure~\ref{dispersion} to Figure~\ref{num} are in fact very well described by the deep water/short wavelength dispersion relation (\ref{dispshort}), as were the waves studied in the experiments~\cite{NJP} and~\cite{Badulin}.

At the blocking lines or horizons discussed in the last Section, the group velocity vanishes and the dispersion curve $\omega(k)$ has a local extremum. Three possible horizons for gravity-capillary waves were identified: the white, blue and negative horizons in Figure~\ref{dispersion}(b). At each horizon two real roots of the dispersion relation coalesce into one double root and then disappear: in the terminology of dynamical systems it is a saddle-node or tangent bifurcation \cite{Nardin}. The order parameter of the bifurcation is the wave number whereas the two control parameters are the velocity $U$ (the ``external field") and the frequency $\omega$ (the ``internal parameter"). Following the approach in~\cite{Nardin}, we find the horizons by looking for a double root $k_2$ of the cubic dispersion relation (\ref{dispshort}):
\begin{equation} \label{double}
(k-k_1)(k-k_2)^2=0,
\end{equation}
where $k_1$ is the remaining simple root. Comparing coefficients of $k$ in (\ref{dispshort}) and (\ref{double}) we obtain expressions for $k_2$ and $k_1$, as follows. The comparison of coefficients gives three equations for the two unknowns $k_1$ and $k_2$; two of these equations are solved for $k_1$ and $k_2$, and the the third equation is then a constraint relating $k_1$ to $k_2$. In the case of positive wave numbers (plus sign in (\ref{dispshort})) this procedure gives
\begin{equation}  \label{kspos}
\fl
\begin{array}{l}
{\displaystyle  k_2=\frac{\rho U^2}{3\gamma} \left( 1\pm \sqrt{1-\frac{3\gamma}{\rho U^4}(g+2\omega U)}\right) }  \\[15pt]
{\displaystyle  k_1=\frac{\rho U^2}{3\gamma} \left( 1\mp 2\sqrt{1-\frac{3\gamma}{\rho U^4}(g+2\omega U)}\right) }
\end{array}
\quad \mbox{(positive wave numbers)}
\end{equation}
with the constraint
\begin{equation}  \label{k1k2}
k_1k_2^2 =\frac{\rho \omega ^2}{\gamma} \quad 
\mbox{(positive wave numbers)}.
\end{equation}
The constraint (\ref{k1k2}), for both sign possibilities in (\ref{kspos}), leads to
\begin{equation}  \label{conpos}
\fl
\omega\left[U^5+\frac{gU^4}{4\omega}+\frac{\gamma \omega ^2 U^3}{\rho g}-\frac{15\gamma \omega U^2}{2\rho }-\frac{6g\gamma  U}{\rho }
-\frac{\gamma g^2 }{\rho \omega}-\frac{27\gamma ^2 \omega ^3 }{4\rho ^2 g}\right]=0 \quad\left[\!\!\! \begin{array}{c} \mbox{positive wave} \\  \mbox{numbers}
\end{array}\!\!\! \right]
\end{equation}
For negative wave numbers  (minus sign in (\ref{dispshort})) $k_2$ and $k_1$ are
\begin{equation}  \label{ksneg}
\fl
\begin{array}{l}
{\displaystyle  k_2=-\frac{\rho U^2}{3\gamma} \left( 1\pm \sqrt{1-\frac{3\gamma}{\rho U^4}(g-2\omega U)}\right) }  \\[15pt]
{\displaystyle  k_1=-\frac{\rho U^2}{3\gamma} \left( 1\mp 2\sqrt{1-\frac{3\gamma}{\rho U^4}(g-2\omega U)}\right) }
\end{array}
\quad \mbox{(negative wave numbers)}
\end{equation}
with the constraint
\begin{equation}
k_1k_2^2 =-\frac{\rho \omega ^2}{\gamma} \quad 
\mbox{(negative wave numbers)}
\end{equation}
that leads to
\begin{equation}  \label{conneg}
\fl
\omega\left[U^5-\frac{gU^4}{4\omega}+\frac{\gamma \omega ^2 U^3}{\rho g}+\frac{15\gamma \omega U^2}{2\rho }-\frac{6g\gamma  U}{\rho }
+\frac{\gamma g^2 }{\rho \omega}+\frac{27\gamma ^2 \omega ^3 }{4\rho ^2 g}\right]=0 \ \left[\!\!\! \begin{array}{c} \mbox{negative wave} \\  \mbox{numbers}
\end{array}\!\!\! \right]
\end{equation}
The significance of the constraints (\ref{conpos}) and (\ref{conneg}) is clear from Figure~\ref{dispersion}(b). With a particular choice of $U$, local extrema in the dispersion plot (corresponding to double roots $k_2$) occur at values of $\omega$ determined by this choice of $U$. For a given $U$, the constraint (\ref{conpos}) or (\ref{conneg}) is a quartic in $\omega$ whose real roots give all the frequencies at which a blocking line (horizon) occurs for this $U$. Alternatively, upon fixing $\omega$ the constraint gives a quintic in $U$ whose real roots are all the flow velocities that give a horizon at this frequency.

\begin{figure}[!htbp]
\begin{center}
\includegraphics[width=12cm]{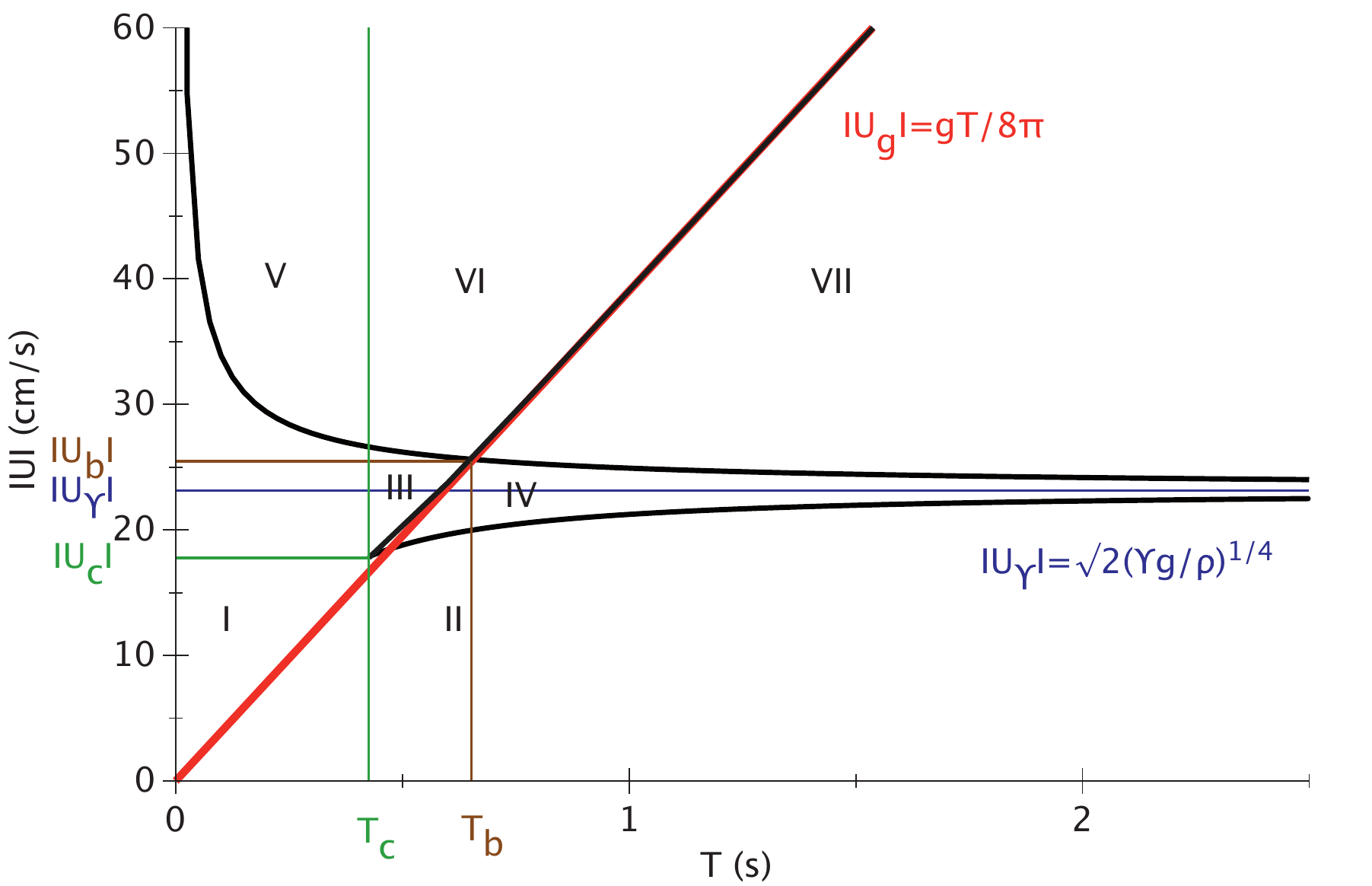}
\caption{The flow speeds $|U|$ at which the white, blue and negative horizons occur, as a function of the wave period $T$ (black curves). The curve lying on or close to the straight red line is the white horizon, the red line showing the pure-gravity white horizon. The curve approaching the asymptote $|U|=|U_\gamma|$ from below is the blue horizon, and the curve approaching this asymptote from above is the negative horizon. }
\label{phase}
\end{center}
\end{figure}

As in the previous Section, we consider only positive $\omega$ and negative $U$, whereas $k$ can be positive or negative (this gives no restriction in the horizon effects). The white horizon in Figure~\ref{dispersion}(b) occurs at a value of $k$ low enough for the influence of the surface tension $\gamma$ in (\ref{dispshort}) to be very small. A very good approximation for the white horizon, which is exact in the pure-gravity case, is therefore obtained by putting $\gamma=0$. Neglecting $\gamma$, the constraint (\ref{conpos}) gives
\begin{equation}  \label{congrav}
\omega U^4 (U+\frac{g}{4\omega}) \simeq 0.
\end{equation}
Note that the approximation (\ref{congrav}) is also obtained in the large $U$ limit, so for large $U$ it becomes the exact constraint at the white horizon for gravity-capillary waves. From (\ref{congrav}) we recover the relation between $U$ and $\omega$ that gives the blocking of gravity waves at a white-hole horizon~\cite{Nardin}
\begin{equation}  \label{Ugrav}
U_g=-\frac{g}{4\omega}=-\frac{gT}{8\pi},
\end{equation}
where $T$ is the period. The corresponding value of the double root $k_2$ is obtained from the $\gamma\to 0$ limit of (\ref{kspos}); the lower sign in the first of (\ref{kspos}) gives the only finite expression:
\begin{equation}   \label{k2grav}
k_2= k_g= \frac{4\omega ^2}{g}=\frac{g}{4U_g^2}
\end{equation}
Equations (\ref{Ugrav})--(\ref{k2grav}) show the exact relationship between the frequency, wave number and counter-flow speed at the white horizon for pure-gravity waves. Note from (\ref{Ugrav}) that the flow speed at the white horizon is proportional to the conserved period of the blocked wave. The straight line $|U_g|$ versus $T$ is shown in red in Figure~\ref{phase}. Superimposed on that red line is a black curve that shows the exact relationship between flow speed and period for gravity-capillary waves. In line with the comments above, the red line agrees very well with the gravity-capillary case except for small $|U|$, and therefore small $T$. The striking feature of the gravity-capillary curve is that it ends at the point labeled $(|U_c|,T_c)$; the white horizon thus does not exist for periods $T$ that are below a critical value $T_c$, or for counter-flows that do not reach a critical speed $|U_c|$. The existence of this threshold can be seen from the dispersion plots in Figure~\ref{rays}: for flow velocity $U_1$ there is no local maximum of the green curve, so no frequency $\omega$ experiences a white horizon at this flow velocity; in contrast, the other flow velocities plotted in the figure all give a local maximum of the green curve and therefore a white horizon for the frequency at this maximum. Similarly, if the period $T$ is too small (frequency $\omega$ too large), the horizontal red line in Figure~\ref{rays} will not intersect a local maximum in the dispersion plot for any $U$, so there can be no white horizon for such periods.

\begin{figure}[!htbp]
\begin{center} 
\includegraphics[width=12cm]{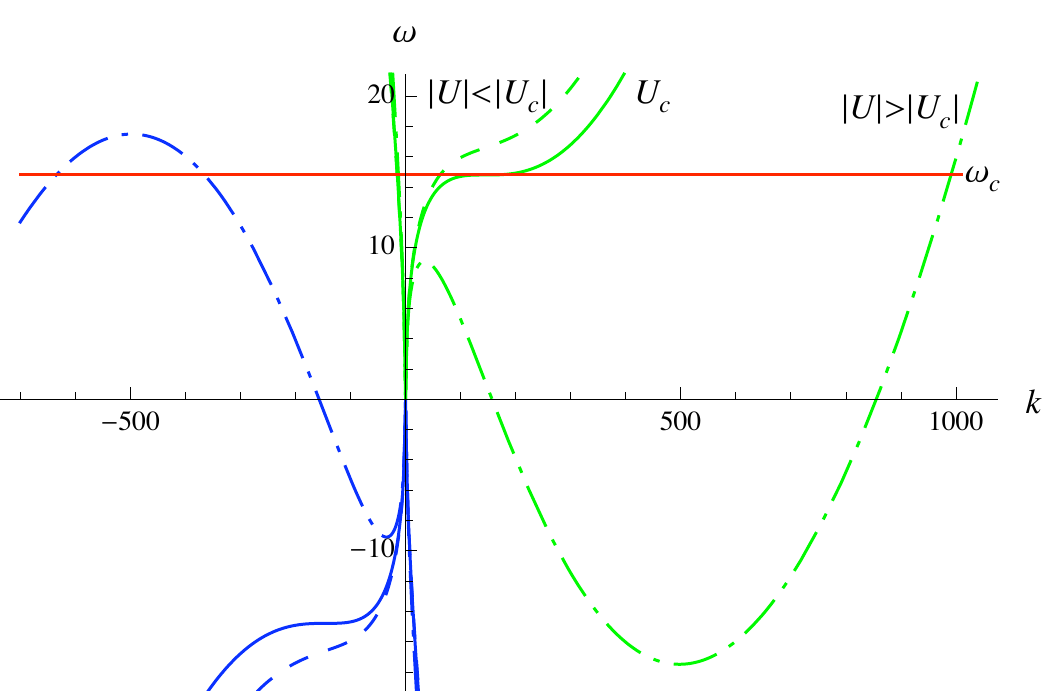} 

\vspace{5mm}

\includegraphics[width=10cm]{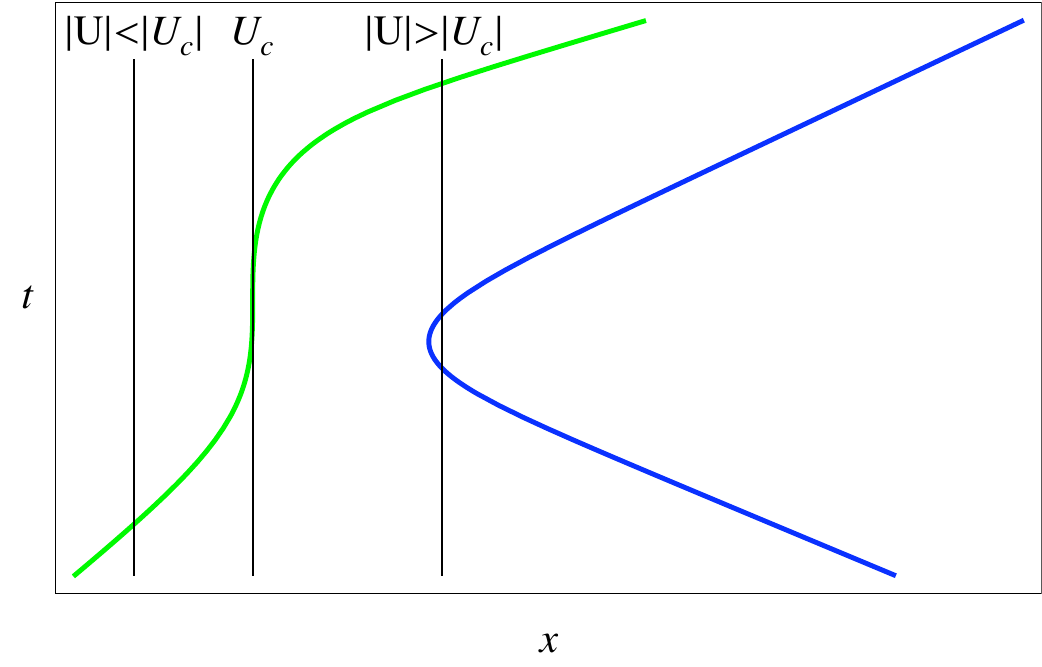}
\caption{Dispersion plots and ray solutions for a wave with period $T_c=2\pi/\omega_c$. The wave with positive co-moving frequency occurs at a point of inflection when $U=U_c$. This means that the corresponding ray (green) has a group velocity that slows to zero when $|U|$ increases to $|U_c|$; the group velocity does not reverse, however, and the ray resumes its propagation into regions of higher $|U|$.  }
\label{inflection}
\end{center}
\end{figure}

Let us look in more detail at the critical values $(|U_c|,T_c)$. Figure~\ref{inflection} shows graphically the occurrence of the threshold for the white horizon. We see that the disappearance of a local maximum in the green curve, as $|U|$ decreases, is accompanied by the disappearance of the local minimum, and at the critical value $U_c$ the two local extrema coalesce to form a point of inflection at frequency $\omega_c$. This shows that $(|U_c|,T_c)$ is also the threshold for the occurrence of the blue horizon, which requires a local minimum in the green curve. At $(|U_c|,T_c)$ the two double roots $k_2$ in (\ref{kspos}) (white and blue horizons) coincide, and in fact the same value is taken by the simple root $k_1$, as can also be seen from Figure~\ref{inflection}. The values $(|U_c|,T_c)$ can be obtained by solving for the point of inflection $\frac{\partial \omega }{\partial k}=\frac{\partial ^2 \omega }{\partial k^2}=0$ in the dispersion relation (\ref{dispshort}) ($k>0$); these two equations can be solved for $U_c$ and the critical wave number $k_c$, and $\omega_c$ then follows from the dispersion realtion. Alternatively, the point of inflection is found by demanding that the square-root expression in (\ref{kspos}) vanishes so that all $k_2$ and $k_1$ coincide. The result is
\begin{equation}  \label{Tc}
T_c=2\pi (3+2\sqrt{3})^{3/4}\left( \frac{\gamma}{\rho g^{3}}\right)^{1/4}=0.425\,\mathrm{s}
\end{equation}
and
\begin{equation}  \label{Uc}
U_c = -\frac{\sqrt{3}}{(3+2\sqrt{3})^{1/4}}\left( \frac{\gamma g}{\rho}\right)^{1/4}=-0.178\,\mathrm{m/s}
\end{equation}
with the wave number
\begin{equation}  \label{kc}
k_c =\frac{1}{(3+2\sqrt{3})^{1/2}}\left(\frac{\rho g}{\gamma}\right)^{1/2}=144\,\mathrm{m}^{-1}.
\end{equation}
In using the method (\ref{double}) of searching for the system parameters at horizons, we noted that they correspond to saddle-node or tangent bifurcations in dynamical-systems theory~\cite{Nardin}.  The cusp ($|U_c|,T_c$) in the $|U|$ vs $T$ diagram (Figure~\ref{phase}) is the point where two saddle-node lines (horizons) intersect. In the terminology of dynamical systems this corresponds to a so-called pitchfork bifurcation~\cite{Poston}.

\begin{figure}[!htbp]
\begin{center} 
\includegraphics[width=12cm]{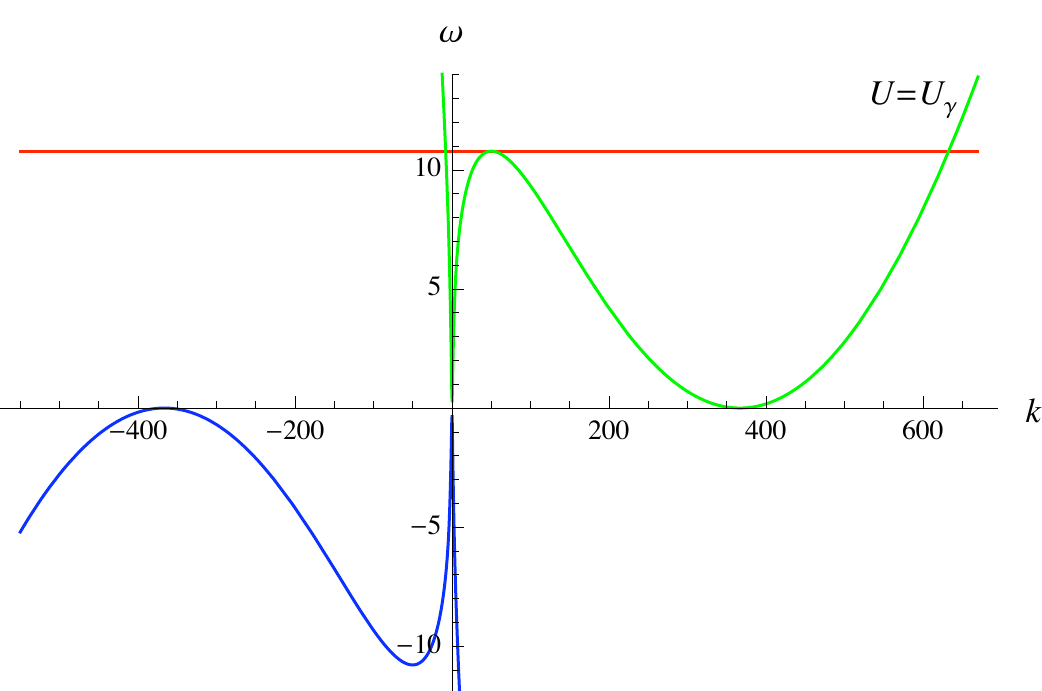} 
\caption{The dispersion plot when the counter-flow velocity is such that the local minimum of the green curve and the local maximum of the blue curve lie on the $k$-axis ($\omega=0$, $T=\infty$). This counter-flow velocity is given by (\ref{Ugamma}) and has the value $U_\gamma=-0.231\,\mathrm{m/s}$. }
\label{bluezero}
\end{center}
\end{figure}

Since $(|U_c|,T_c)$ is also a threshold for the existence of the blue horizon, the curve in the $|U|$ versus $T$ diagram (Figure~\ref{phase}) relating the period to the flow speed at the blue horizon must also end at $(|U_c|,T_c)$; this curve is also shown in Figure~\ref{phase} and it is seen to approach an asymptotic value, labelled $U_\gamma$, as $T\to\infty$. This is because $T\to\infty$ means $\omega\to 0$ and the dispersion plot in Figure~\ref{bluezero} shows that the local minimum giving the blue horizon occurs at $\omega=0$ for a finite non-zero $U$ that we call $U_\gamma$. We find the flow velocity $U_\gamma$ as follows. The constraint (\ref{conpos}) relates the values of $U$ and $\omega$ at all horizons (double roots of the dispersion relation) for waves with positive $k$. Hence by taking $\omega\to 0$ in (\ref{conpos}) we obtain $U_\gamma$; this limit of (\ref{conpos}) gives
\begin{equation}
\frac{gU^4}{4}-\frac{\gamma g^2}{\rho }=0,
\end{equation}
so
\begin{equation} \label{Ugamma}
U_\gamma=-\sqrt{2}\left(\frac{\gamma g}{\rho}\right)^{1/4}=-0.231\,\mathrm{m/s}.
\end{equation}
The wave vector at the local minimum (blue horizon at $\omega=0$) in  Figure~\ref{bluezero} is found by inserting the velocity $U_\gamma$ into the expression for the double root $k_2$ in (\ref{kspos}) and taking the upper sign: 
\begin{equation}  \label{k2zeroblue}
k_\gamma=\left(\frac{\rho g}{\gamma}\right)^{1/2}.
\end{equation}
The other local extremum at $k>0$ in Figure~\ref{bluezero}, the white horizon for a non-zero $\omega$ given by the horizontal red line in the figure, occurs at
\begin{equation} \label{kwhitezeroblue}
k_2=0.137\left(\frac{\rho g}{\gamma}\right)^{1/2},
\end{equation}
and the corresponding $\omega$ (red line) is
\begin{equation} \label{omwhitezeroblue}
\omega=0.180\left(\frac{\rho g^3}{\gamma}\right)^{1/4},
\end{equation}
where the exact but lengthy numerical coefficients have not been reproduced. The simple root $k_1$ for $U=U_\gamma$ is zero for $\omega=0$, while for  $\omega$ given by (\ref{omwhitezeroblue}) (red line Figure~\ref{bluezero}) the simple root is (intersection of red line with green curve at large $k>0$ in  Figure~\ref{bluezero})
\begin{equation} 
k_1=1.73\left(\frac{\rho g}{\gamma}\right)^{1/2}.
\end{equation}
These last three results are obtained from the constraint (\ref{conpos}) and the expressions (\ref{kspos}) for the double and single roots, with $U=U_\gamma$.
 
We see from Figure~\ref{bluezero} that in the limit $\omega\to 0$ ($T\to\infty$) the negative horizon also occurs at the flow velocity $U_\gamma$, as well as the blue horizon, and the wave vector at the negative horizon is minus that at the blue horizon, $-k_\gamma$. Figure~\ref{bluezero} shows that, for waves with positive laboratory frequency $\omega$, the counter-flow velocity $U_\gamma$ is the threshold for the existence of waves with negative co-moving frequency; the threshold flow velocity for such waves is the threshold for the negative horizon. It follows that the curve in the $|U|$ versus $T$ diagram (Figure~\ref{phase}) relating the period to the flow speed at the negative horizon must lie above the line $|U|=|U_\gamma|$ and asymptotically approach this line as $T\to\infty$. This negative-horizon curve is also plotted in Figure~\ref{phase}; it lies above the blue-horizon curve but shares with it the asymptote $|U|=|U_\gamma|$. Unlike the white- and blue-horizon curves in Figure~\ref{phase}, which both end at the cusp $(|U_c|,T_c)$ for small $T$, the negative-horizon curve diverges to $|U|\to\infty$ as $T\to 0$. This behaviour of the negative horizon is clear from the dispersion plots because as  $T\to 0$ ($\omega\to\infty$) the flow speed $|U|$ must increase without limit in order for the local maximum in the negative-$k$ curve to reach the horizontal frequency line (see for example Figure~\ref{rays}).

The flow velocity $U_\gamma$ that appears as an asymptote in the $|U|$ versus $T$ diagram (Figure~\ref{phase}) has additional significance in fluid mechanics. Firstly, it is a well-known property of gravity-capillary waves on {\it static} water ($U=0$) that the minimum phase velocity of the waves is given (apart from the sign) by the expression (\ref{Ugamma}) for $U_\gamma$. The velocity $U_\gamma$ is also important in the case of {\it shear flows}, i.e.\ velocity profiles that change with the fluid depth. It was shown by Caponi {\it et al.}~\cite{Caponi} that a sufficient condition for a shear flow to become spontaneously unstable is for the flow velocity on the surface of the fluid to exceed $U_\gamma$; the instability leads to the generation of gravity-capillary waves on the fluid surface~\cite{Caponi}. Another example in shear flows is the appearance of negative-energy waves at the interface of two fluid layers, which occurs when the relative velocity of the layers exceeds $U_\gamma$; this is related to the famous Kelvin-Helmholtz instability, as discussed by Fabrikant and Stepanyants~\cite{FS}.

\begin{figure}[!htbp]
\begin{center} 
\includegraphics[width=12cm]{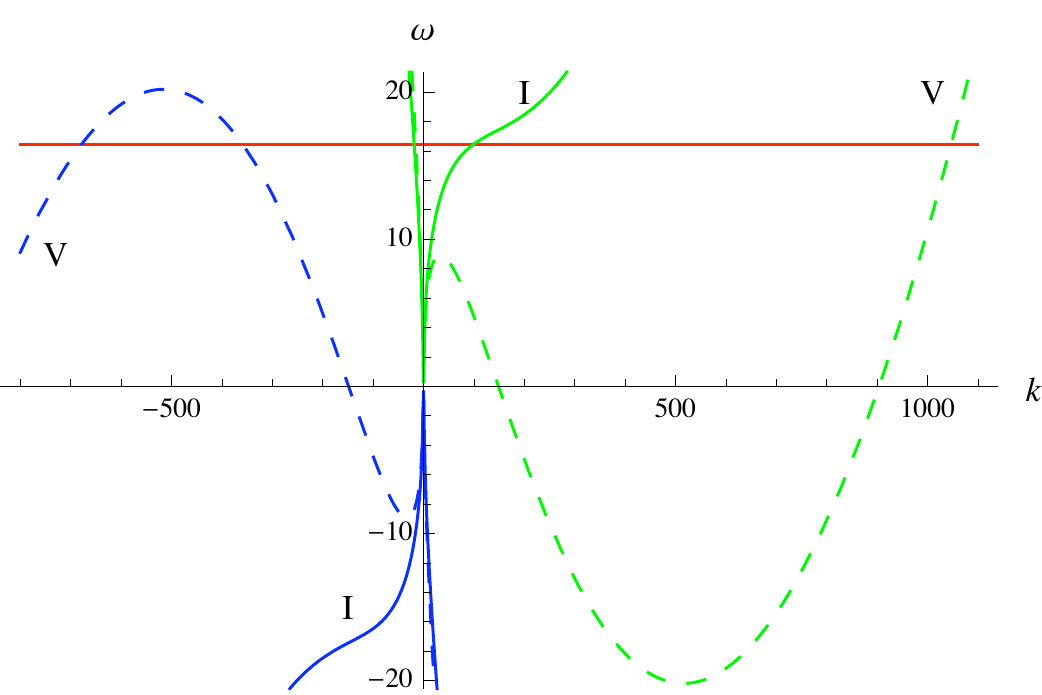} 

\vspace{5mm}

\includegraphics[width=10cm]{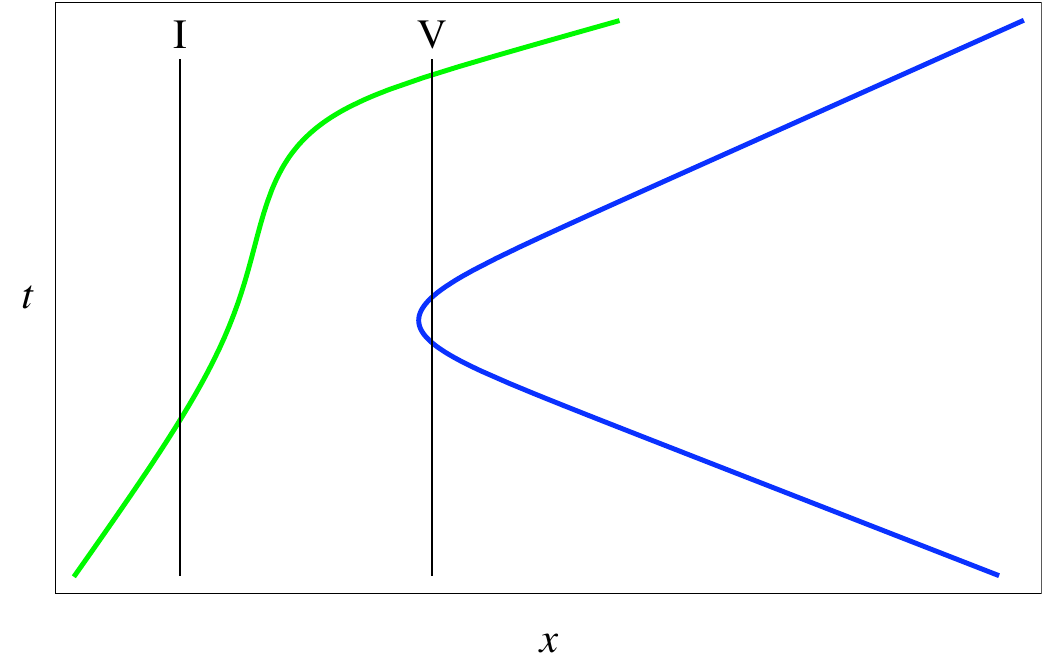}
\caption{Dispersion plots and ray solutions for waves with period $T<T_c$. The Roman numerals I and V refer to counter-flow speeds that lie in the regions labeled by these numerals in Figure~\ref{phase}. The period is $T=0.382\,\mathrm{s}$ and the flow velocities are $-0.159\,\mathrm{m/s}$ (I) and  $-0.277\,\mathrm{m/s}$ (V).}
\label{ItoV}
\end{center}
\end{figure}

Figure~\ref{phase} allows the classification of the behaviour of gravity-capillary waves on a stationary counter-flow. The period $T$ is conserved in the wave evolution so by fixing a vertical line in the figure one can distinguish five qualitatively different possibilities for waves of a single frequency:

1. $T<T_c$. As one moves into regions in the velocity profile $U(x)<0$ with higher counter-flow speeds, one moves from region I in Figure~\ref{phase} into region V. The dispersion plots and ray solutions for this case are shown in Figure~\ref{ItoV}. Notable is the fact that there is no white or blue horizon.

2.  $T=T_c$. Here the line of constant $T$ is the vertical green line in Figure~\ref{phase} that separates region I from regions II and III and passes through the cusp point $(|U_c|,T_c)$. The dispersion and ray plots are shown in Figure~\ref{inflection}; as already discussed, this case is the threshold for the appearance of the white and blue horizons.

\begin{figure}[!htbp]
\begin{center} 
\includegraphics[width=12cm]{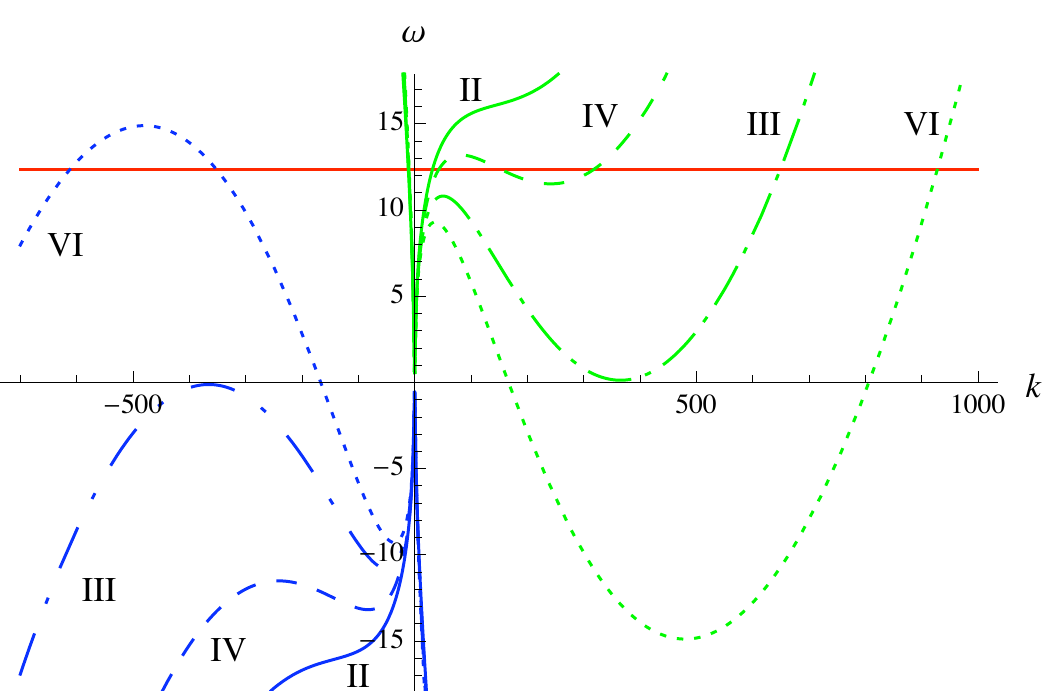} 

\vspace{5mm}

\includegraphics[width=10cm]{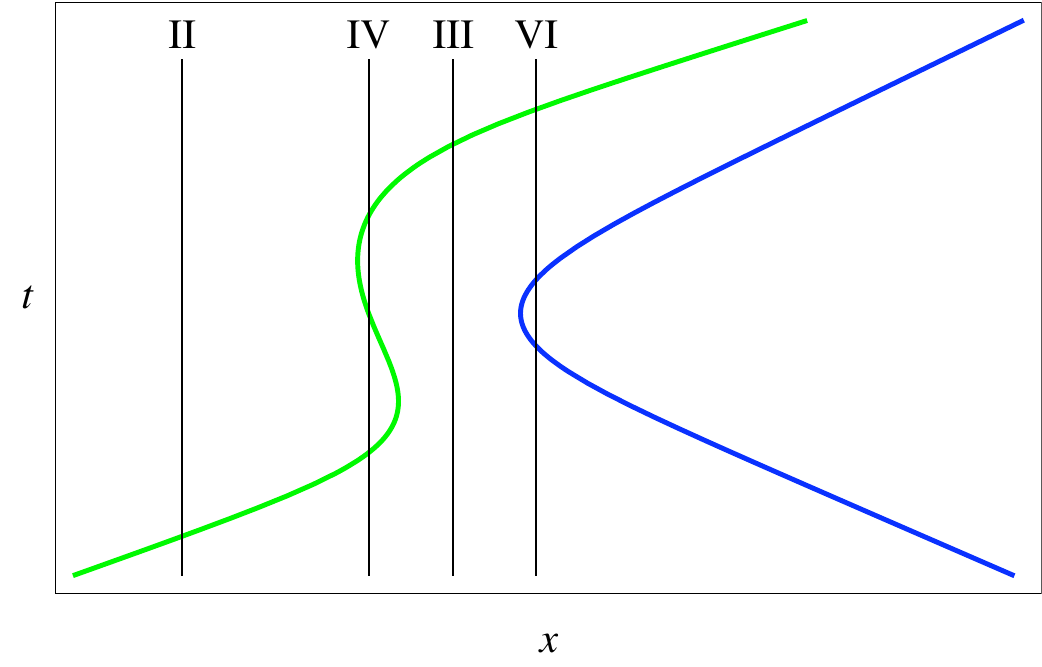}
\caption{Dispersion plots and ray solutions for waves with period $T_c<T<T_b$. The Roman numerals refer to counter-flow speeds that lie in the regions labeled by these numerals in Figure~\ref{phase}. The period is $T=0.510\,\mathrm{s}$ and the flow velocities are $-0.169\,\mathrm{m/s}$ (II),  $-0.194\,\mathrm{m/s}$ (IV), $-0.231\,\mathrm{m/s}$ (III) and $-0.267\,\mathrm{m/s}$ (VI).}
\label{IItoIVtoIIItoVI}
\end{center}
\end{figure}

3. $T_c<T<T_b$. The line of constant $T$ lies between the green and brown vertical lines in Figure~\ref{phase} and so passes through region III. Here increasing counter-flow speeds takes us from region II to IV to III to VI in Figure~\ref{phase}. The dispersion plots and ray solutions for this case are shown in Figure~\ref{IItoIVtoIIItoVI}. Here there is a white and blue horizon, and the white horizon occurs at a lower counter-flow speed than the negative horizon.

\begin{figure}[!htbp]
\begin{center} 
\includegraphics[width=12cm]{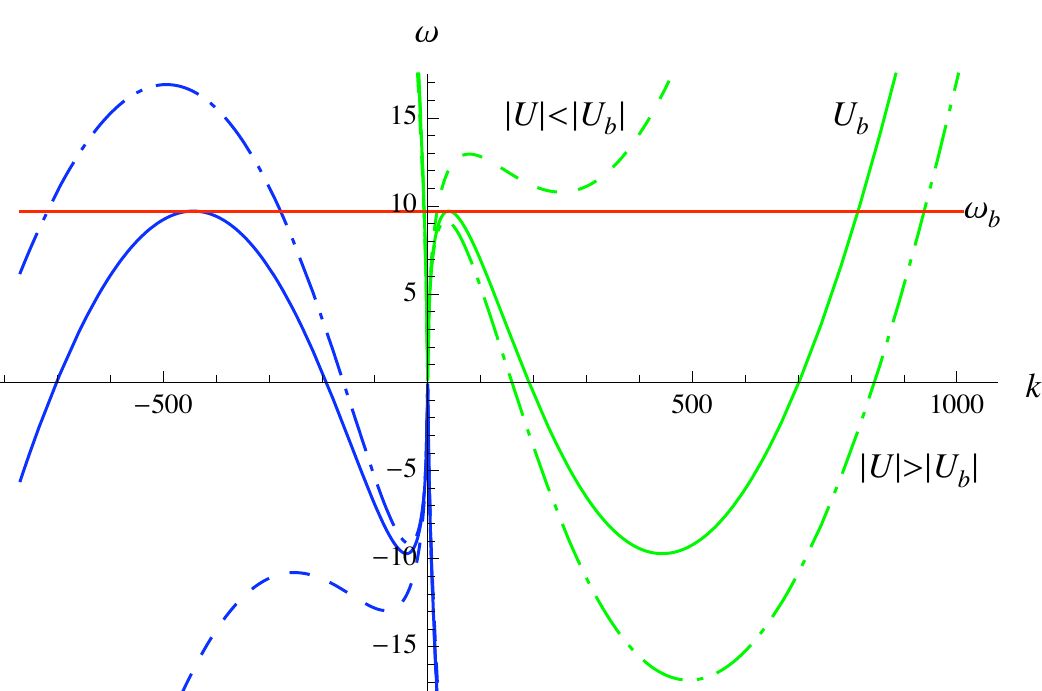} 

\vspace{5mm}

\includegraphics[width=10cm]{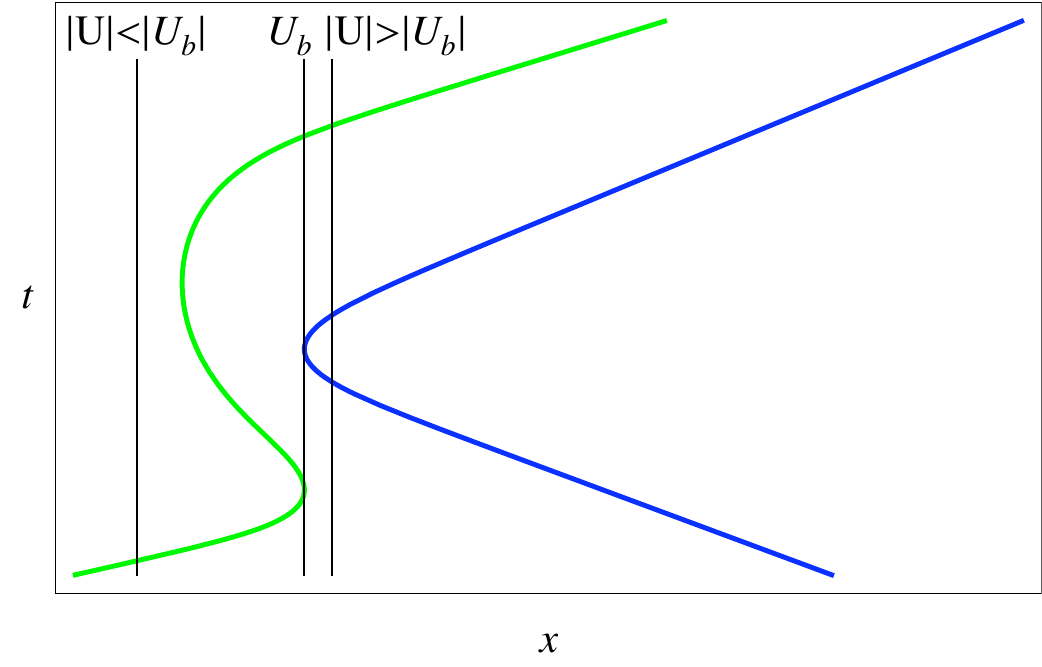}
\caption{Dispersion plots and ray solutions for waves with period $T=T_b$ (see Figure~\ref{phase}). Both the white and negative horizons occur at the same counter-flow velocity $U=U_b$, the point $(T_b,|U_b|)$ being the intersection of the curves for these two horizons in Figure~\ref{phase}. For water, $T_b=0.647\,\mathrm{s}$, $U_b=-0.255\,\mathrm{m/s}$.}
\label{dispb}
\end{center}
\end{figure}

4. $T=T_b$, defined by the brown vertical line in Figure~\ref{phase}. The line $T=T_b$ passes through the point where regions III, IV, VII and VI meet, at $|U|=|U_b|$. The significance of the point $(|U_b|,T_b)$ is that it is the intersection of the white-horizon curve and the negative-horizon curve; this means that for a wave with period $T_b$ the white horizon occurs at the same counter-flow speed as the negative horizon. Figure~\ref{dispb} confirms this in the dispersion plots and ray solutions. The values $(U_b,T_b)$ must be found numerically and for water they are $T_b=0.647\,\mathrm{s}$, $U_b=-0.255\,\mathrm{m/s}$.

\begin{figure}[!htbp]
\begin{center} 
\includegraphics[width=12cm]{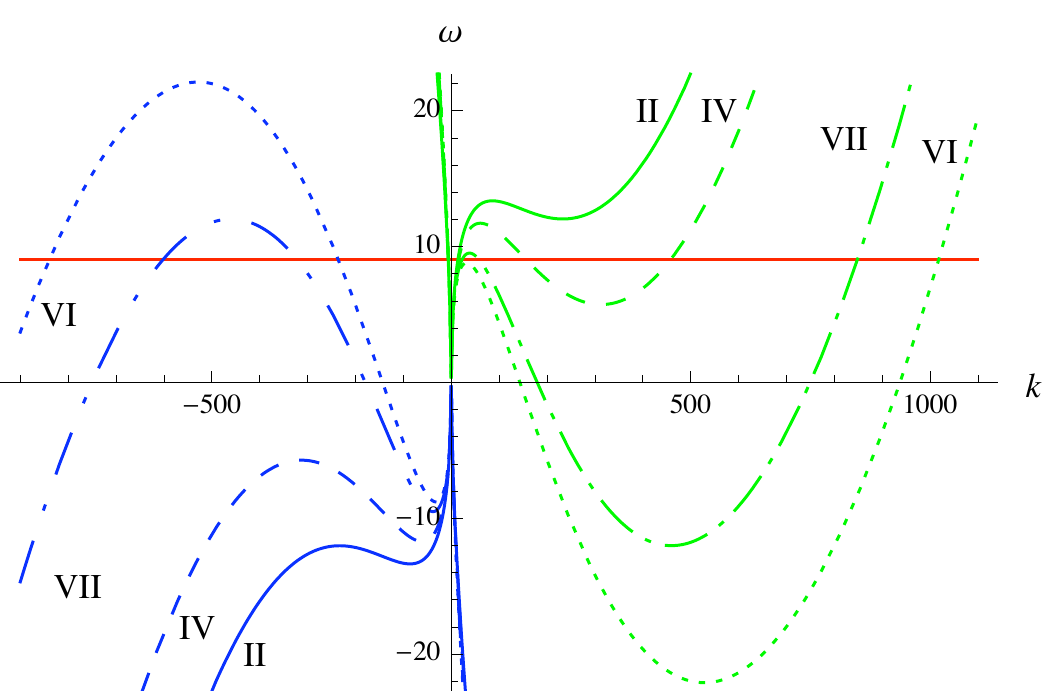} 

\vspace{5mm}

\includegraphics[width=10cm]{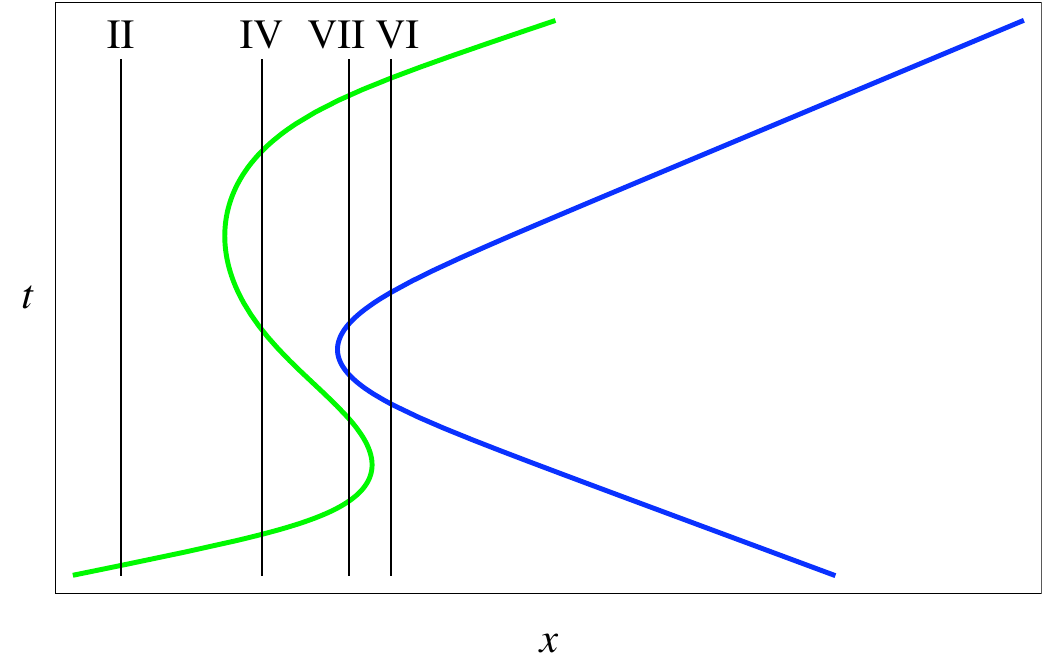}
\caption{Dispersion plots and ray solutions for waves with period $T>T_b$. The Roman numerals refer to counter-flow speeds that lie in the regions labeled by these numerals in Figure~\ref{phase}. The period is $T=0.692\,\mathrm{s}$ and the flow velocities are $-0.192\,\mathrm{m/s}$ (II),  $-0.214\,\mathrm{m/s}$ (IV), $-0.260\,\mathrm{m/s}$ (VII) and $-0.281\,\mathrm{m/s}$ (VI).}
\label{IItoIVtoVIItoVI}
\end{center}
\end{figure}

5.  $T>T_b$. Here the line of constant $T$ passes through region VII in Figure~\ref{phase}. Increasing counter-flow speeds takes us from region II to IV to VII to VI. The dispersion plots and ray solutions for this case are shown in Figure~\ref{IItoIVtoVIItoVI}. Here there is a white and blue horizon, and the white horizon occurs at a higher counter-flow speed than the negative horizon (compare carefully with case~3 above).

Trulsen~\cite{Trulsen} inferred from his results the structure of the $(|U|,T)$-diagram of Figure~\ref{phase}, but without the negative-horizon curve. He also derived the existence of the cusp $(|U_c|,T_c)$ as a triple-root solution of the dispersion relation~\cite{Trulsen}. 

Recently, we observed (with continuous waves trains) the regions VI, VII and IV of Figure~\ref{phase},  as reported in our experimental $|U|$ versus $T$ diagram~\cite{NJP}. The distinction between the regions II and IV was unclear from our data. Our wave-maker was limited to a minimum period of $0.5\,\mathrm{s}$ which is higher than $T_c=0.4255\,\mathrm{s}$. Our focus was on the conversion into waves with negative co-moving frequency rather than conversion into capillary waves in the double-bouncing behaviour described above. Conversion to capillary waves is difficult to observe experimentally because of the rapid dissipation of capillary waves (we were unaware of the work by Badulin {\it et al.}~\cite{Badulin}). In addition, we used a rather high period (far from $T_c$) to get long wavelengths of the ingoing waves since the waves with negative co-moving frequency should be produced with a drastic reduction of the wavelength according to the dispersion relation. We were surprised to find indications of waves with negative co-moving frequency even without wave blocking (a white horizon).

Badulin {\it et al.}~\cite{Badulin} performed experiments with wave packets (three to ten wave cycles centred on periods in the range $T=0.33$--$0.66\,\mathrm{s}$) sent on a counter-flow with speed $|U|$ between $0.04$ and $0.3\,\mathrm{m/s}$ over a sloping bottom. Double bouncing of the input waves was observed with a strong reduction in both wavelength (from $\lambda=0.2\,\mathrm{m}$ to $2\,\mathrm{mm}$!) and amplitude. These authors presented only one photograph, at $T=0.52\,\mathrm{s}$, of the conversion phenomenon and one measurement of the amplitude of waves as a function of the position/velocity ($T=0.5\,\mathrm{s}$ and $\rmd U/\rmd x=0.1\,\mathrm{s}^{-1}$), but beautiful measurements of the effect of the velocity on the wavelength. No results for periods less than $T_c=0.425\,\mathrm{s}$ were reported.

\section{A thermodynamic analogy}
In the previous Section we summarized the behaviour of gravity-capillary waves (with $kh\gg 1$) on a stationary counter-flow by means of a diagram in the $(|U|,T)$-parameter space (Figure~\ref{phase}). This diagram allows one to visualize the evolution of an incident wave of a single frequency (which is conserved), as was illustrated in the five cases in Figures~\ref{ItoV}, \ref{inflection}, \ref{IItoIVtoIIItoVI},\ref{dispb} and~\ref{IItoIVtoVIItoVI}. In this Section we note a similarity between Figure~\ref{phase} and a phase diagram in thermodynamics, where the horizon lines in Figure~\ref{phase} are analogous to the lines separating different phases (first-order phase transition). In particular, the cusp point $(|U_c|,T_c)$ in Figure~\ref{phase} looks like a critical point (second-order phase transition) in a phase diagram.

\begin{figure}[!htbp]
\begin{center}
\includegraphics[width=12cm]{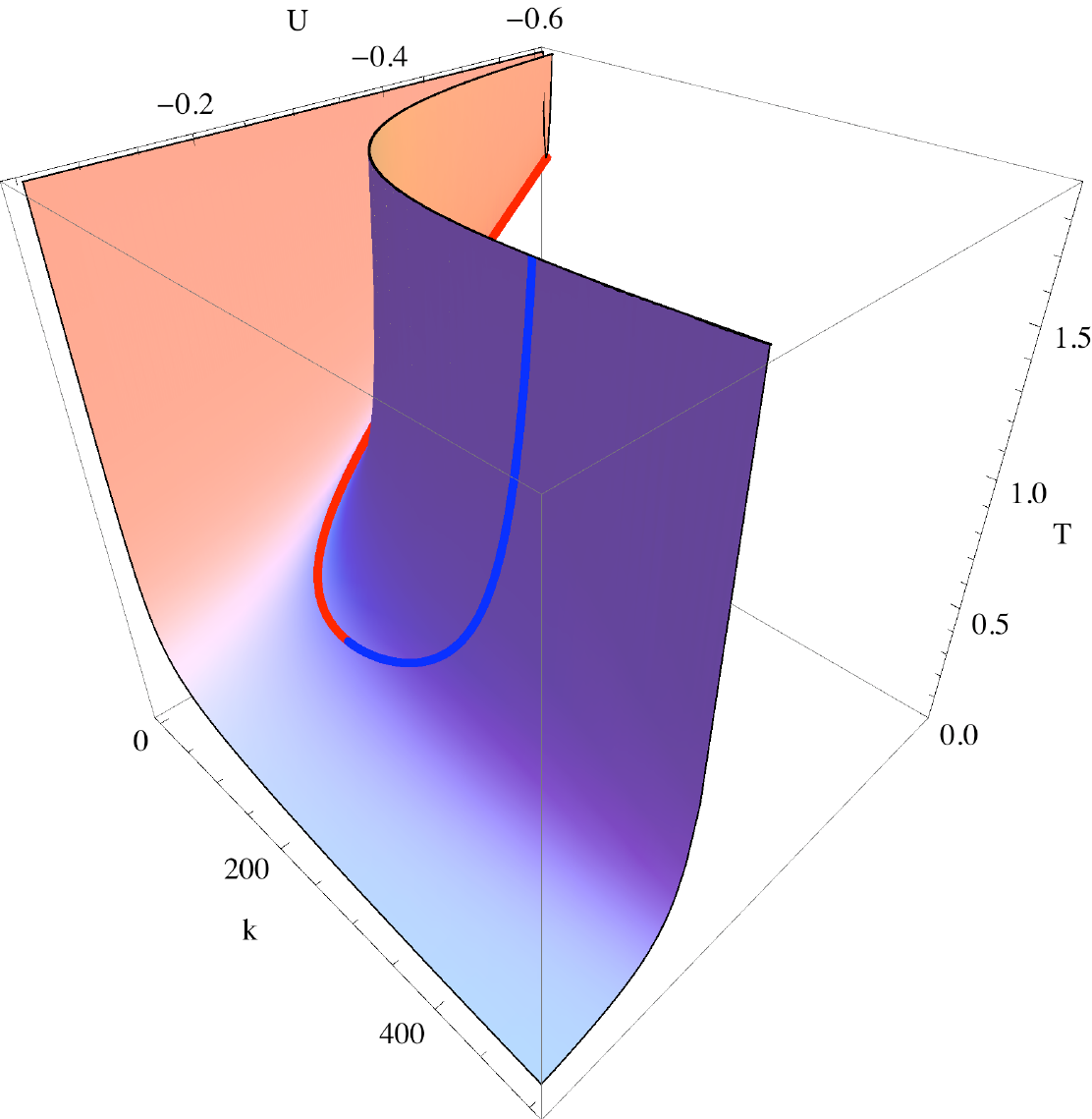}
\caption{The state of a wave on a counter-flow is a surface in $(U,k,T)$-space given by (\ref{UkTsurface}). This surface has a fold catastrophe in the positive-$k$ sector, similar to the fold catastrophe of the van der Waals gas~\cite{Poston}. The red curve on the state surface is the white horizon and the blue curve is the blue horizon. The horizon curves meet at $(U_c,k_c,T_c)$, which is analogous to the critical point of the van der Waals gas. The white- and blue-horizon curves in Figure~\ref{phase} are the projection of the curves on the state surface to the $(U,T)$-plane. }
\label{cata}
\end{center}
\end{figure}

Let us explore this thermodynamic analogy a little further. In thermodynamics a system is described by an equation of state of the form $f(P,V,\Theta)=0$, where $P$ is the pressure, $V$ is the volume and $\Theta$ is the temperature. For example, the equation of state of a van der Waals gas can be written~\cite{Poston}:
\begin{equation} \label{vdW}
V^3-\left(nb+\frac{nR\Theta}{P} \right)V^2+\frac{n^2 a}{P}V-\frac{n^3ab}{P}=0,
\end{equation}
where $n$ is the number of molecules divided by Avogadro's number, $R$ is the gas constant, $b$ relates to the non-zero volume of the molecules and $a$ is a measure of the molecular interaction. A familiar property of the van der Waals gas is the existence of a critical point in the $(P,\Theta)$-phase diagram; this is associated with a fold catastrophe in the surface (\ref{vdW}) in $(P,V,\Theta)$-space that constitutes the state-space of the gas~\cite{Poston}. Now the dispersion relation (\ref{dispshort}), written as
\begin{equation}  \label{UkTsurface}
k^3 -\frac{\rho U^2}{\gamma}k^2+\frac{\rho}{\gamma}\left(g+\frac{4\pi U}{T}\right)k-\frac{4\pi^2\rho}{\gamma T^2}=0,
\end{equation}
describes the state of a wave on a counter-flow as a surface in $(U,k,T)$-space and this surface has the same kind of fold catastrophe as the van der Waals gas---see Figure~\ref{cata}. The connection between wave blocking and catastrophe theory was inferred a long time ago by Peregrine and Smith~\cite{PS} and more recently by Trulsen~\cite{Trulsen}. The white and blue horizons appear as curves on the state surface in Figure~\ref{cata}, which converge and join at the ``critical point" $(U_c,k_c,T_c)$ given by (\ref{Tc})--(\ref{kc}). Only the positive-$k$ part of the state surface is shown in Figure~\ref{cata}, since this contains the point analogous to the thermodynamic critical point. By projecting the surface in Figure~\ref{cata} on to the $(U,T)$-plane one obtains the part of the ``phase diagram" Figure~\ref{phase} containing the white and blue horizons, similar to the $(P,\Theta)$-diagram of the van der Waals gas~\cite{Poston}. 

We can summarize the analogy between the van der Waals gas and waves on a counter-flow by the following table, which we stress describes only a qualitative relationship:

\begin{center}
\begin{tabular}{|l|c|r|}
  \hline
   \hline
  Van der Waals Gas & Wave-Current Interaction  \\
   \hline
    \hline
    Volume $V$ & Wave Number $k$   \\
   \hline
   Temperature $\Theta$ & Frequency $\omega $    \\
   \hline
   Pressure $P$ & Flow Velocity $U$  \\
   \hline
  Compressibility $\frac{\partial V}{\partial P}$ &  Susceptibility $
\frac{\partial k }{\partial U}$  \\
   \hline
  Spinodal Line &  Blocking Line  \\
    \hline
   Perfect Gas $P=\frac{Nk_B \Theta}{V} $ &  Pure Advection $U=
\frac{\omega }{k}$   \\
   \hline
\end{tabular}
\end{center}
The perfect gas is seen to correspond to very large $\omega$ and $U$ (pure advection $U=\omega/k$ of the surface waves). In addition, the perfect gas is obtained by setting the parameters $a$ and $b$ to zero; similarly, pure advection of the surface waves corresponds to setting the parameters $\gamma$ and $g$ to zero. 

\begin{figure}[!htbp]
\begin{center}
\includegraphics[width=15cm]{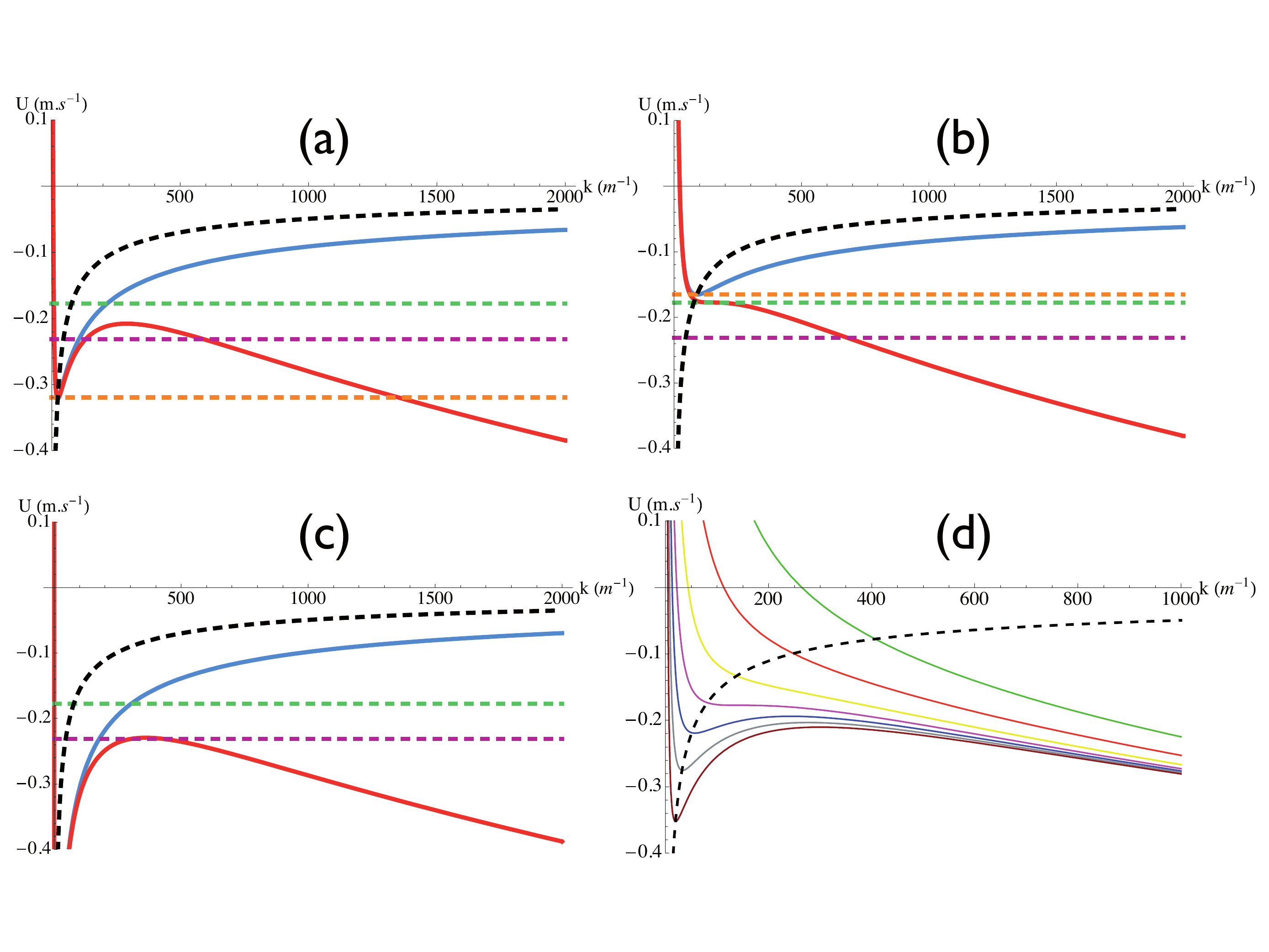}
\caption{The red curves in Figures~(a), (b) and (c) show $U$ as a function of $k$ for different periods $T$: Figure (a) is for $T>T_c$,  (b) for $T=T_c$ and (c) for $T\to \infty$. Figure (d) shows a series of plots of $U(k)$ for different $T$; the green curve is for $T\to 0$ (pure advection). The blue curves in Figures (a)--(c) show the pure-gravity case. The black dotted line in all four figures shows, in the pure-gravity case, the locus of the local minimum of $U(k)$ as $T$ changes. The horizontal dotted purple line is $U_\gamma$; this is the value of $U$ at the local maximum of $U(k)$ for  $T\to \infty$ (Figure (c)). The horizontal dotted green line is $U_c$; this  is the value of $U$ at the point of inflection of $U(k)$ for  $T=T_c$ (Figure (b)). The horizontal dotted orange line is $U_g$, which is proportional to $T$ (see (\ref{Ugrav})); it gives the value of $U$ at the local minimum in the pure gravity case (blue curves in Figures (a)--(c)). }
\label{Uvsk}
\end{center}
\end{figure}

The fold catastrophe in Figure~\ref{cata} can also be projected to the $(U,k)$-plane (Figure~\ref{Uvsk}). From the dispersion relation (\ref{dispshort}) these projections are given by
\begin{equation}  \label{Uk}
U(k)=\frac{\omega}{k} -\sqrt{\frac{g}{k}+\frac{\gamma}{\rho}k}
\end{equation}
with a fixed value of $\omega$ ($T$). Figure~\ref{Uvsk} shows the curves $U(k)$ for different periods $T$, for both the gravity-capillary case (red lines in (a), (b) and (c)) and the pure-gravity case (blue lines in (a), (b) and (c)). Figure~\ref{Uvsk}(d) shows $U(k)$ for a range of periods $T$ in the gravity-capillary case; these isoperiod curves are analogous to the Andrews isotherms for a real gas. 

We have seen from dispersion plots that wave-blocking corresponds to local extrema of the function $\omega (k)$. It follows from the implicit function theorem that the minimum (maximum) of $U(k)$ corresponds to the maximum (minimum) of $\omega (k)$, and therefore blocking lines are given by the local extrema of $U(k)$. These local extrema $\partial U/\partial k=0$ are the analogues of the spinodal line $\partial P/\partial V=0$ of the van der Waals gas. Without surface tension (blue lines in Figure~\ref{Uvsk}(a)--(c)), $U(k)$ has a single minimum that lies on the dotted black line in these plots. With surface tension (red lines in Figure~\ref{Uvsk}(a)--(c)), $U(k)$ has in addition a local maximum. At $T=T_c$ (Figure~\ref{Uvsk}(b)) the minimum and the maximum of $U(k)$ merge and an inflection point appears; this corresponds to the cusp in the $(U,T)$-plane (Figure~\ref{phase}). We can define a mechanical susceptibility $\chi _m =\left(\frac{\partial k}{\partial U}\right)_T$, analogous to the isothermal compressibility coefficient $\chi _\theta =-\frac{1}{V}\left(\frac{\partial V}{\partial P}\right)_\theta$, that diverges at the horizons, just as the compressibility of the gas diverges at the spinodal line.

In the limit of infinite period $T \to \infty$ ($\omega \to 0$) we found that the wave is described by a flow speed (\ref{Ugamma}) and wave number (\ref{k2zeroblue}). We recover this result from the $\omega=0$ case of (\ref{Uk}), which is
\begin{equation}  \label{UkTinf}
U(k)= -\sqrt{\frac{g}{k}+\frac{\gamma}{\rho}k}.
\end{equation}
This is plotted in Figure~\ref{Uvsk}(c) and has a maximum $\frac{\partial U}{\partial k}=0$ at wave number
\begin{equation}
k=\sqrt{\frac{\rho g}{\gamma}}=\frac{1}{l_c}=k_\gamma ,
\end{equation}
where $l_c$ is the capillary length. This wave number in (\ref{UkTinf}) reproduces $U_\gamma$, given by (\ref{Ugamma}).

The principle of corresponding states implies that the properties of real gas are universal functions of the state variables scaled to the critical point. For the Van der Waals gas, it is well known that the equation of state can be written in a universal form around the critical point ($V_c=3nb$, $\theta _c=\frac{8a}{27bR}$ and $P_c=\frac{a}{27b^2}$):
\begin{equation}
\left(P_r+\frac{3}{V_r^2}\right)\left(3V_r-1\right)=8\theta _r,
\end{equation}
where the subscript $r$ means reduced variable ($V_r=\frac{V}{V_c}$, $\theta_r=\frac{\theta}{\theta _c}$ and $P_r=\frac{P}{P_c}$). Similarly, using the scalings ($k_r=\frac{\sqrt{\gamma }k}{\sqrt{\rho g}}$, $\omega_r=\frac{\gamma^{1/4}\omega}{(\rho g^3)^{1/4}}$ and $U_r=\frac{\rho ^{1/4}U}{(\gamma g)^{1/4}}$), we find the universal dispersion relation:
\begin{equation}
\left(\omega _r -U_r k_r\right)^2=k_r (1+k_r^2).
\end{equation}
Whatever the fluid (surface tension, density), its wave-like behavior will be the same close to the cusp. The dimensionless form of the constraint (\ref{conpos}) becomes:
\begin{equation}
\omega _r U_r^5+\frac{1}{4}U_r^4+\omega _r^3U_r^3-\frac{15}{2}\omega _r^2U_r^2-6\omega _rU_r-1-\frac{27}{4}\omega _r^4=0.
\end{equation}
One recovers $U_\gamma =-\sqrt{2}\left(\frac{\gamma g}{\rho}\right)^{1/4}$ by imposing $\omega _r=0$. By introducing another scaling $U'_r=U_r \omega _r= \frac{U\omega}{g}$ the constraint reads:
\begin{equation}
{U'}_r^5+\frac{1}{4}{U'}_r^4+\omega _r^4\left({U'}_r^3-\frac{15}{2}{U'}_r^2-6U'_r-1-\frac{27}{4}\omega _r^4\right)=0
\end{equation}
One recovers $U_g=-\frac{1}{4}\frac{g}{\omega}$ by imposing $\omega _r=0$.

\section{Conclusions and Perspectives}
We have described the interaction of linear gravity-capillary waves with a counter-flow, with emphasis on the various horizon effects (wave blocking in fluid-mechanics terminology). The case of waves with negative co-moving frequency has been included throughout; these waves are crucial for the Hawking effect and they have been neglected in the fluid-mechanics literature on wave blocking. The Hawking effect is a remarkable process in which an incident wave generates a wave with negative co-moving frequency, with a resulting \emph{amplification} of the incident wave (this implies an extraction of energy from the flow). It has been shown that this process is robust in the presence of dispersion~\cite{Unruh95,Brout,Corley,Jacobson99,Himemoto,Saida,Unruh05,SU08} and the linear theory of surface waves falls into the class of systems that exhibit the Hawking effect~\cite{SU}. Experimental evidence of the generation of waves with negative co-moving frequency was reported in~\cite{NJP} and further experiments are planned.

We presented analytical results for the deep water/short wavelength case $kh\gg 1$ that are more comprehensive than those given elsewhere. A similarity of the state space of the waves to that of a thermodynamic system was pointed out; the curves in the state space representing the horizon lines are analogous to curves separating thermodynamic phases and there is even an analogue of a thermodynamic critical point. 

Sch\"utzhold and Unruh showed that the regime of gravity-wave propagation in an effective Schwarzschild-like metric corresponds to the shallow water limit $kh\ll 1$~\cite{SU}. The interaction with the white-hole horizon necessarily tunes a wave out of this regime ($kh\ll1$) into the $kh\gg 1$ regime in which dispersion causes the effective-metric description to break down. In other words, for pure gravity waves the white horizon is not dispersive when $kh\ll 1$ whereas it is dispersive when $kh\gg 1$. On the other hand, waves in the $kh\gg 1$ regime considered in this paper stay in this regime in the interaction with the counter-flow.

Previous results on the shape of waves at a blocking line (horizon) can be developed further for gravity-capillary waves. The horizons for surface waves can be treated as examples of saddle-node lines, which also describe caustics in optics. It is well known that an Airy function describes both the intensity of light close to an optical caustic of the fold type~\cite{Berry,Adam} and the water shape for the amplitude of gravity waves close to a blocking line~\cite{Smith,PS,Nardin}. It will be shown elsewhere that the Airy function depends on a  ``stopping length" $L_s$ (roughly the width of the arch of the Airy function) which scales like $L_s \approx gT^{5/3}\left(\frac{dU}{dx}\right)_{x=x^*}^{-1/3}$ where $x^*$ is the position of the white horizon, and $\left(\frac{dU}{dx}\right)_{x=x^*}$ is the ``surface gravity" at the horizon. An experimental measurement of the Airy shape has been carried out by Chawla and Kirby~\cite{Chawla1}. If surface tension is taken into account, we have seen that there is a critical point at the intersection of two saddle-node lines (white and blue horizons). We anticipate that the wave at the critical point will be described by a Pearcey catastrophe integral~\cite{PS,Trulsen} due to the superposition of two Airy catastrophe integrals for the two saddle-node lines.

\ack
We are indebted to Yury Stepanyants, Erwann Aubry and Gil Jannes for fruitful discussions. This research was supported by the R\'egion PACA (Projet exploratoire HYDRO), the Conseil G\'en\'eral 06, the Scottish Government, the Royal Society of Edinburgh and the Royal Society of London.


\section*{References}
\begin{thebibliography}{99}

\bibitem{Unruh}
Unruh W G 1981 {\it Phys. Rev. Lett.}  {\bf 46} 1351

\bibitem{Novello}
Novello M, Visser M and Volovik G E (eds) 2002
{\it Artificial black holes}
(Singapore: World Scientific)

\bibitem{Volovik}
Volovik G E 2003
{\it The Universe in a Helium Droplet} 
(Oxford: Oxford University Press)

\bibitem{BLV}
Barcelo C, Liberati S and Visser M 2005 {\it Living Rev. Rel.} {\bf 8} 12

\bibitem{SUbook}
Sch\"utzhold R and Unruh W G (eds) 2007
{\it Quantum Analogues: 
From Phase Transitions to Black Holes and Cosmology}
(Berlin: Springer)

\bibitem{Fiber}
Philbin T G, Kuklewicz C, Robertson S, Hill S, 
K\"onig F and Leonhardt U 2008 {\it Science} {\bf 319} 1367

\bibitem{tHooft}
'tHooft G 1985
{\it Nucl. Phys.} B {\bf 256} 727

\bibitem{Jacobson}
Jacobson T 1991
{\it Phys. Rev.} D {\bf 44} 1731

\bibitem{Hawking}
Hawking S W 1975
{\it Commun. Math. Phys.} {\bf 43} 199

\bibitem{Unruh95}
Unruh W G 1995 {\it Phys. Rev.} D {\bf 51} 2827

\bibitem{Brout}
Brout R, Massar S, Parentani R and Spindel P 1995
{\it Phys. Rev.} D {\bf 52} 4559

\bibitem{Corley}
Corley S 1998
{\it Phys. Rev.} D {\bf 57} 6280

\bibitem{Jacobson99}
Jacobson T and Mattingly D 1999
{\it Phys. Rev.} D {\bf 61} 024017

\bibitem{Himemoto}
Himemoto Y and Tanaka T 2000
{\it Phys. Rev.} D {\bf 61} 064004

\bibitem{Saida}
Saida H and Sakagami 2000
{\it Phys. Rev.} D {\bf 61} 084023

\bibitem{Unruh05}
Unruh W G and Sch\"utzhold R 2005 {\it Phys. Rev.} D {\bf 71} 024028

\bibitem{SU08}
Sch\"utzhold R and Unruh W G 2008 {\it Phys. Rev.} D  {\bf 78} 041504(R)

\bibitem{SU}
Sch\"utzhold R and Unruh W G 2002 {\it Phys. Rev.} D  {\bf 66} 044019

\bibitem{BHlaser}
Corley S and Jacobson T 1999
{\it Phys. Rev.} D {\bf 59} 124011

\bibitem{Barcelo}
Barcel\'{o} C, Cano A, Garay, L J and Gannes G 2006 {\it Phys. Rev.} D {\bf 74} 024008

\bibitem{visser}
Visser M and Weinfurtner S 2007 {\it PoS(QG-Ph)} 042 (\emph{Preprint} arXiv:0712.0427[gr-qc])

\bibitem{NJP}
Rousseaux G, Mathis C, Maissa P, Philbin T G and Leonhardt U 2008 {\it New J. Phys.} {\bf 10} 053015

\bibitem{FS}
Fabrikant A L and Stepanyants Y A 1998 {\it Propagation of Waves in Shear Flows} (Singapore: World Scientific)

\bibitem{Huang}
Huang H 2008 {\it Chinese Sci. Bull.} {\bf 53} 3267

\bibitem{Peregrine}
Peregrine D H 1976 {\it Adv. Appl. Mech.} {\bf 16} 9

\bibitem{Smith}
Smith R 1976 {\it J. Fluid Mech.} {\bf 77} 417

\bibitem{Chawla1}
Chawla A and Kirby J T 1998 {\it Proc. 26th Int. Conf. on Coastal Engineering} (Copenhagen: ASCE) p~759

\bibitem{Chawla2}
Chawla A and Kirby J T 2002 {\it J. Geophys. Res.} {\bf 107} 10.1029/2001JC001042

\bibitem{Suastika}
Suastika I K 2004 Wave blocking (PhD thesis, Technische Universiteit Delft, The Netherlands)

\bibitem{Igor}
Lavrenov I 2003 {\it Wind-Waves in Oceans} (Berlin: Springer)

\bibitem{Choi}
Choi W 2009 {\it Mathematics and Computers in Simulation} {\bf 80} 29

\bibitem{Schaffer}
Schaffer S 1979 {\it J. Hist. Astron.}  {\bf 10} 42

\bibitem{Darrigol}
Darrigol O 2005 {\it Worlds of Flow: A History of Hydrodynamics from the Bernoullis to Prandtl} (Oxford: Oxford University Press)

\bibitem{Badulin}
Badulin, S I, Pokazeev K V and Rozenberg A D 1983 {\it Izvestiya Atmos. Ocean. Phys.} {\bf 19} 1035

\bibitem{Dingemans}
Dingemans M W 1997 {\it Water Wave Propagation Over Uneven Bottoms} (Singapore: World Scientific)

\bibitem{PS}
Peregrine D H and Smith R 1979 {\it Philos. Trans. R. Soc. London} A {\bf 292} 341

\bibitem{Basovich}
Basovich A Y and Talanov V I 1977 {\it lzv. Akad. Nauk SSSR, Fiz. Atmos. Okeana} {\bf 13} 766

\bibitem{Unruh76}
Unruh W G 1976 {\it Phys. Rev.} D {\bf 14} 870

\bibitem{Baschek}
Baschek B 2005 Wave-current interaction in tidal fronts {\it 14th Aha Hulikoa Winter Workshop: Rogue Waves} (Honolulu, Hawaii)

\bibitem{sti79}
Stiassnie M and Dagan G 1979 {\it J. Fluid Mech.} {\bf 92} 119

\bibitem{Nardin}
Nardin J-C,  Rousseaux G and Coullet P 2009 {\it Phys. Rev. Lett.} {\bf 102} 124504

\bibitem{shy90}
Shyu J-H and Phillips O M 1990 {\it J. Fluid Mech.} {\bf 217} 115

\bibitem{shy99}
Shyu J-H and Tung C-C 1999 {\it J. Fluid Mech.} {\bf 396} 143

\bibitem{Trulsen}
Trulsen K 1995 The influence of currents, long waves and wind on gravity-capilary waves (PhD thesis, Massachusetts Institute of Technology)

\bibitem{TM}
Trulsen K and Mei C C 1993 {\it J. Fluid Mech.} {\bf 251} 239

\bibitem{Klinke}
Klinke J and Long S R 2000 Generation of short waves by wave-current interaction {\it Proc. Geoscience and Remote Sensing Symposium, IEEE International Publication} {\bf 3} 1084

\bibitem{Poston}
Poston T and Stewart I 1998 {\it Catastrophe Theory and Its Applications} (New York: Dover)

\bibitem{Caponi}
Caponi E A, Yuen H C, Milinazzo F A and Saffman P G 1991 {\it J. Fluid Mech.} {\bf 222} 207

\bibitem{Berry} 
Berry M V 1981 {\it Les Houches Lecture Series Session XXXV} (Amsterdam: North-Holland) p 453

\bibitem{Adam}
Adam J A 2002 {\it Phys. Rep.} {\bf 356} 229

\end{thebibliography}
\end{document}